\documentclass[%
aip,
amsmath,amssymb,
reprint,%
]{revtex4-1}

\usepackage{graphicx}
\usepackage{dcolumn}
\usepackage{bm}

\usepackage[utf8]{inputenc}
\usepackage[T1]{fontenc}
\usepackage{mathptmx}

\begin{document}

\preprint{AIP/123-QED}

\title[ ]{Fat Tails and Black Swans: Exact Results for Multiplicative Processes with Resets}
 
\author{D. H. Zanette}
 \email{zanette@cab.cnea.gov.ar}
\affiliation{ 
Centro At\'omico Bariloche and Instituto Balseiro, Comisi\'on Nacional de Energ\'{\i}a At\'omica and Universidad Nacional de Cuyo, Consejo Nacional de Investigaciones Cient\'{\i}ficas y T\'ecnicas, Av. E. Bustillo 9500, R8402AGP San Carlos de Bariloche, R\'{\i}o Negro, Argentina. 
}

\author{S. Manrubia}
\email{smanrubia@cnb.csic.es}
 \affiliation{
Grupo Interdisciplinar de Sistemas Complejos (GISC), Madrid, Spain; Centro Nacional de Biotecnolog\'{\i}a (CSIC); c/ Darwin 3, 28049 Madrid, Spain.
}

\date{\today} 

\begin{abstract}
We consider a class of  multiplicative  processes which, added with stochastic reset events, give origin to stationary distributions with power-law tails ---ubiquitous in the statistics of social, economic, and ecological systems. Our main goal is to provide a series of exact results on the dynamics and asymptotic behaviour of increasingly complex versions of a basic multiplicative process with resets, including discrete and continuous-time variants and several degrees of randomness in the parameters that control the process. In particular, we show how the power-law distributions are built up as time elapses, how their moments behave with time, and how their stationary profiles become quantitatively determined by those parameters. Our discussion emphasizes the connection with financial systems, but these stochastic processes are also expected to be fruitful in  modeling   a wide variety of social and biological phenomena.  
\end{abstract}

\maketitle

\begin{quotation}
City sizes, word usage, surname abundance, personal income or stock market returns are examples of power-law distributed quantities. Such kind of distribution, ubiquitous in the natural and social sciences, holds ``atypical'' properties that have awoken the interest of researchers for over a century.\cite{auerbach:1913,estoup:1916,yule:1924,zipf:1949,simon:1955} The dynamics of this kind of data exhibit extreme, catastrophic, ``unexpected'' black-swan-like events.\cite{taleb:2007}  Distribution moments, such as the average or the variance, are highly volatile and poorly predict future properties of the process. In that context of uncertainty, knowledge of the dominant mechanisms underlying power-law distributions is relevant to directly compare the short-time properties of data series to actual asymptotic properties, and to eventually evaluate the reliability of forecast algorithms. Stochastic multiplicative processes (SMPs) with reset events, introduced two decades ago\cite{manrubia:1999} as a generic mechanism to generate power laws, have multiple applications in a variety of situations.\cite{evans:2011,montero:2013} In this contribution, we derive several finite-time properties of SMPs with reset events with the aim of improving our understanding of the poor predictability of the dynamical process. The discussion of our results in a financial context clarifies the relationship between gain and risk in investing strategies, and provides clues to control the frequency and magnitude of extreme events.
\end{quotation}

\section{\label{sec1}Introduction}

\begin{quote}
``At first glance, the facts of human life do not seem to subordinate themselves, as do the phenomena of nature, to certain general laws. Statistics show, however, that this is just a gradual difference due to the more complicated nature of human relations, and that man only has to learn to read the statistics correctly in order to extract more general conclusions; and it is not uncommon to find interesting and strange laws.''
\end{quote}

The quote above opens a report written in 1913 by Felix Auerbach,\cite{auerbach:1913} where he discussed the skewed abundance distribution of cities of various sizes in Germany, in what is likely the first published document on the topic. Beyond the identification of an uncommon statistical pattern, Auerbach's   paper reveals a visionary intuition  on the existence of general laws  underlying the statistical properties of social phenomena. Three decades later, in his book ``Human Behavior and The principle of least effort,''\cite{zipf:1949} George Kingsley Zipf compiled multiple data  with the aim of demonstrating regularities and subsequently discussing plausible, simple underlying principles. The distribution of city sizes and the frequency of words in written texts  ---an observation that goes back to 1916\cite{estoup:1916}--- are  undoubtedly the two best-known examples, but Zipf also analyzed the length of intervals between repetitions in a Mozart concerto, the number of retail stores of like kind, the number of passengers travelling by airway, the personal income in different countries  ---first described by Vilfredo Pareto in 1896\cite{pareto:1896}--- and the number of composers of chamber music as a function of their year of birth, among many others. He  ordered each set of events by decreasing size, $S(k)$, assigning rank $k=1$ to the largest event, $k=2$ to the second largest, and so on. Though the relationship $S(k) \propto k^{-\alpha}$, with $\alpha\approx 1$,  was pervasively found (and became known as Zipf's law), there were also many cases of data characterized by  different values of $\alpha$, and even by  other functional relations. Ever since, research on the mechanisms underlying emergent properties of collective human behaviour has been pursued with much effort and also remarkable success.\cite{newman:2004} 

Quantities whose frequency distributions are given by power-law functions have always awaken especial interest in the study of a variety of social systems, as it has the search for simple processes behind this kind of distribution.\cite{MM,zanette:2007} Pure stochastic multiplicative processes (SMPs) yield lognormal distributions for the relevant variables,\cite{montroll:1982} and can produce {\it bona fide} power laws when acting in conjunction with additional mechanisms. Actually, SMPs appear at the core of many successful explanatory models for power laws, such as Yule's birth-death process to explain the distribution of taxa,\cite{yule:1924} or Simon's model to derive the abundance of words in written texts.\cite{simon:1955} These early models rationalize power-law distributions in other systems as well, as in city sizes\cite{zanette:2018} or growth of business firms.\cite{fu:2005} Furthermore, suitable variations of Simon's model reproduce the observed abundance of family names,\cite{zanette:2001,manrubia:2002} while it yields distributions that fit remarkably well the actual frequency of words in written texts\cite{zanette:2005} and the usage of notes in musical compositions.~\cite{zanette:2006}  

Stochastic processes with multiplicative noise plus reinjection  are characterized by distributions with power-law tails.\cite{takayasu:1997,sornette:1998PRE} This kind of mechanism has applications, for instance, in population dynamics and investment portfolio growth.\cite{sornette:1998PhysA} A variant of reinjection that sets a minimum value for the dynamical variable~\cite{levy:1996} can be interpreted in an economic context as a subsidy that keeps individuals above a critical poverty line. SMPs with conservation of the total population, which is a generalization of Zeldovich's intermittency model,\cite{zeldovich:1987} has been used to explain Zipf's law for cities.\cite{zanette:1997} Introduction of reset events in SMPs renders power laws with exponents which depend on the reset probability and on the distribution of growth rates,\cite{manrubia:1999} a situation that can account for the power-law tail of the personal income distribution.\cite{nirei:2004} Though reset events were introduced in the generic context of SMPs yielding power laws,\cite{manrubia:1999} this mechanism has been subsequently applied to a broad variety of problems.\cite{evans:2011,montero:2013,montero:2017,evans:2019} 

SMPs belong to a class of  processes that can exhibit non-self-averaging effects. In physics, this property was first described for spin glasses, where the fact that different realizations of the process visit different areas of the phase space, even in the thermodynamic limit, yields different observable quantities (e.g., moments of the distribution of visited states) for each of the system replicas.\cite{mezard:1987} Later, it was shown that  non-self-averaging behavior was present in various systems,\cite{derrida:1997,derrida:1999} such as  sums of power-law distributed random variables, branching processes, and one-dimensional random walks with return to the origin ---a simple case of stochastic additive processes with resets.\cite{meilahn:2015} A major consequence of the lack of self-averaging is that quantitative properties of the system estimated through averages over time or realizations can be highly unreliable and deeply differ from the  actual asymptotic properties of the stochastic process. 

\begin{figure}[ht]
\centering
\includegraphics[width=\columnwidth]{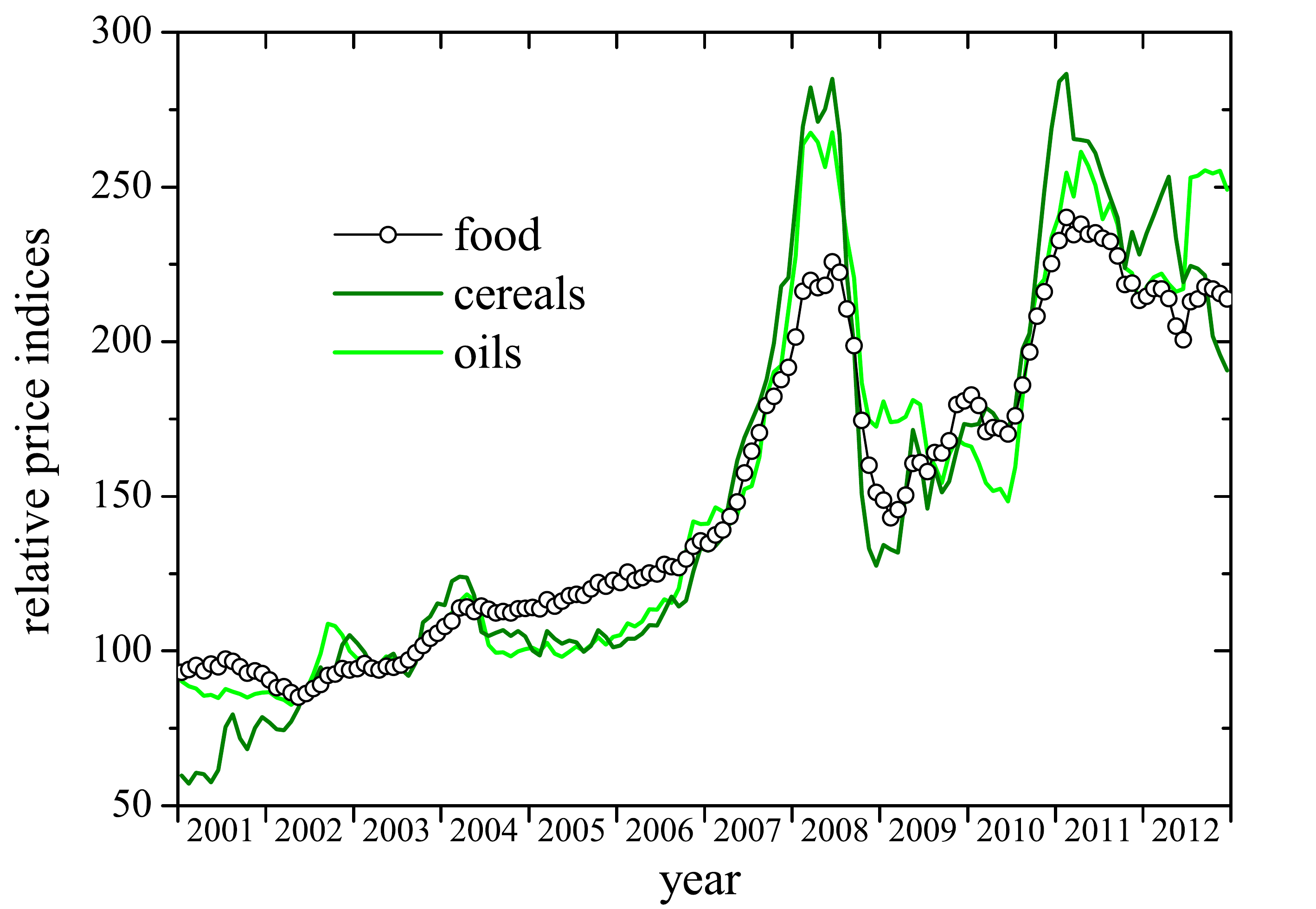}
\caption{Monthly price indices for three major commodities between years 2001 and 2012 (with 2002-2004 average $=100$). Sustained multiplicative growth might precede severe drops to low values. Among the factors put forward {\it a posteriori} to explain such events, one finds weather, increasing demand for meat, use of biofuels, and variations in currency exchange rates. However, on the basis of a deterministic model, it has been shown that the two sharp peaks around 2008 and 2011 are specifically due to investor speculation.\cite{lagi:2011} Public-domain data from the Food and Agriculture Organization (United Nations, www.fao.org).}
\label{fig:PriceIndex}
\end{figure}

Interestingly, the same phenomenology has been observed in economics and finance, where its relevance is difficult to overstate.\cite{gabaix:2009} In these contexts, multiplicative processes are implicit in Gibrat's law,\cite{gibrat:1931} (or ``law or proportionate effect''), which states that the rate of growth of a business firm is independent of its absolute size. Though, as a pure SMP, Gibrat's law implies a lognormal distribution of the relevant variable, the ample evidence of power-law distributed quantities in economics and finance,\cite{gabaix:2009} suggests that additional mechanisms must be at play. Indeed, since sustained exponential growth is unfeasible in reality, Gibrat's law cannot hold at all times. Financial crises and market crashes act as ``control mechanisms'' in the form of catastrophic, punctuated interruptions of the idealized multiplicative growth. Actually, examples of such events abound: price indices suffered severe drops in the crises of 2008 and 2011\cite{lagi:2011} (see Fig.~\ref{fig:PriceIndex}); in the  1980s, some banks lost more money than they ever made in their history;\cite{caprio:1999} financial bubbles more than often cause market crashes, like the well-known Wall Street crash of 1929 and many others. The abrupt loss of  gains in a time much shorter than that required to  accumulate them is commonplace in finance, and resets appear as a  qualitatively suitable mechanism to mimic such dynamics.

Nassim Taleb's  Black Swan Theory\cite{taleb:2007AS} explains the prominent role of such  ``unexpected'' catastrophic events ---as rare black swans among ordinary birds--- not only in finance, but also in many other contexts, such as history, technology, and science. These hardly predictable, large-magnitude occurrences have vastly stronger effects than regular episodes, due to psychological biases and a generalized poor  understanding of the role of probability in social phenomena, related to the subjective notions of luck and fate.\cite{taleb:2007} Such disproportionate consequences are a direct corollary of the unavoidable shortness of historical records, which tends to magnify the exceptionality of those events.     

In this contribution, we revisit SMPs with resets with the aim of improving our understanding of the dynamics of such processes. We derive a number of exact results that allow for precise comparison with quantities obtained from dynamical simulations and clarify the limitations of time-series or realization averages to estimate the moments of the stochastic process ---and therefore to achieve statistical predictions of future events based on knowledge of the past. Sections \ref{sec2}, \ref{sec3}, and \ref{sec4} discuss three models of increasing complexity and present exact results for their dynamics. Their applicability in a financial context is further discussed in Section \ref{sec5}, which closes this paper by highlighting other possible applications and future extensions of SMPs with resets. 

\section{\label{sec2}Uniform multiplication and reset}

The simplest version of the class of multiplicative processes that we consider here is a Markov chain for a variable $x_t$, evolving in discrete time $t=0,1,\dots$, with $x_t>0$ for all $t$ and $x_0=1$. At each time step, $x_t$ is either multiplied by a constant positive factor $\mu \neq 1$ or it is reset to its initial value. The two instances occur with probabilities $1-r$ and $r$, respectively ($0<r<1$). Namely, 
\begin{equation} \label{ev-n}
    x_{t+1}=\left\{
    \begin{array}{ll}
    \mu x_t    & \mbox{with probability $1-r$},  \\
       1  &   \mbox{with probability $r$}.
    \end{array}\right.
\end{equation}
Through the change of variables $x'= | \ln x /\ln \mu |$, this stochastic process is equivalent to the so-called Sisyphus random walk, for which several of the results discussed in this section have been obtained in previous work.\cite{montero:2016}  For the sake of concreteness, we focus on the choice $\mu>1$, which implies that $x_t \ge 1$ for all $t$. Power-law tails in the probability distribution for large $x$, in fact, develop for $\mu>1$ only.  However, results can be straightforwardly extended to the case $\mu<1$, by exploiting the symmetry of the problem under the transformation $\mu\to \mu^{-1}$, $x_t\to x_t^{-1}$. In this case, the probability exhibits power-law behavior for $x\to 0$.  

\begin{figure}[ht]
\centering
\includegraphics[width=\columnwidth]{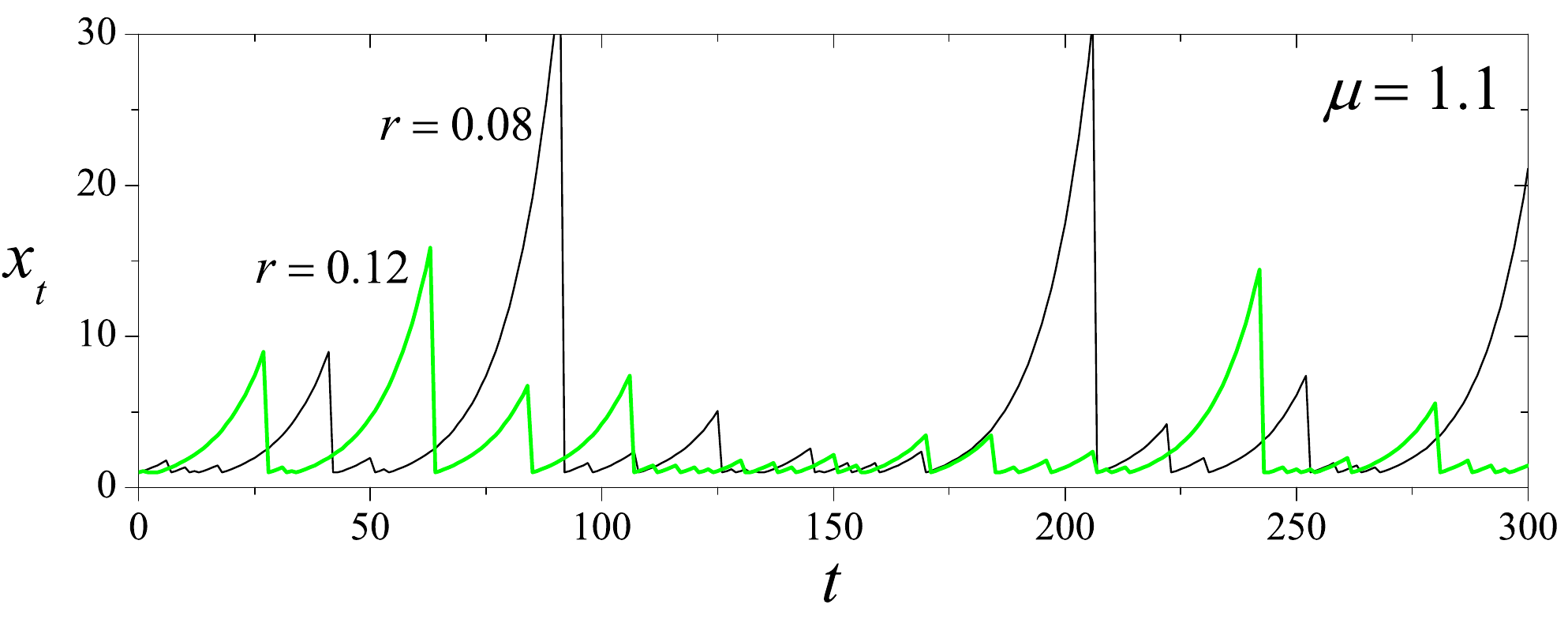}
\caption{Two typical realizations of the stochastic process of Eq.~(\ref{ev-n}), for $\mu=1.1$ and two values of the reset probability: $r=0.08$ (dark curve) and $r=0.12$ (light-shaded curve).}
\label{fig1}
\end{figure}

Figure \ref{fig1} shows the time dependence of $x_t$ along two realizations of   process (\ref{ev-n}), with $\mu=1.1$ and different values of the reset probability $r$. The evolution consists of a succession of ``bursts'' of exponential growth induced by the multiplicative process, each of them terminated by a reset event. For $r=0.08$, the less frequent reset events occasionally allow for rather high bursts, as compared with those obtained for $r=0.12$ in the time span of the simulation. 

The stochastic process (\ref{ev-n}) can be readily dealt with, by noting that the variable $x_t$ can only adopt the values $1,\mu,\mu^2, \dots$ or, generally, $ \mu^m$, for $m=0,1,2,\dots$.  From Eq.~(\ref{ev-n}) it immediately follows that the probability $p_t^{(m)}$ that the variable equals  $\mu^m$ satisfies the Chapman-Kolmogorov equation
\begin{equation}
p_{t+1}^{(m)}    = (1-r) p_t^{(m-1)} (1-\delta_{m0}) + r\delta_{m0},
\end{equation}
where $\delta_{ij}$ is Kronecker's delta.  The solution reads
\begin{equation} \label{disc1}
    p_t^{(m)}=\left\{
    \begin{array}{ll}
     r(1-r)^m  &   \mbox{for $m<t$},   \\
    p_0^{(m-t)} (1-r)^t   & \mbox{for $m\ge t$},
    \end{array}\right.
\end{equation}
in terms of a generic initial condition $p_0^{(m)}$. For $x_0=1$, we have $p_0^{(m)}=\delta_{m0}$, and
\begin{equation}  \label{16}
    p_t^{(m)}=\left\{
    \begin{array}{ll}
     r(1-r)^m  &   \mbox{for $m<t$},   \\
     (1-r)^m  &   \mbox{for $m=t$}, \\
     0   & \mbox{for $m>t$},
    \end{array}\right.
\end{equation}
cf.~Eq.~(7) in Ref.~39. In Eqs.~(\ref{disc1}) and (\ref{16}), all the information about the initial condition is accounted for in the range $m \ge t$ so that, as time elapses, it becomes progressively relegated to exponentially higher values of $x$. The long-time stationary behavior of the probability $p_t^{(m)}$ is built up from small values of $m$,  given by the first line in both equations: 
\begin{equation} \label{stat0}
p_\infty^{(m)} = r (1-r)^m,
\end{equation}
cf.~Eq.~(11) in Ref.~39.

The moment $\langle x^\gamma \rangle_t =\sum_m \mu^{\gamma m} p_t^{(m)}$  can be exactly calculated for any order $\gamma$ at all times. It reads
\begin{equation} \label{mom-n}
  \langle x^\gamma \rangle_t =\frac{r }{1-(1-r)  \mu^\gamma }  + \left[\langle x^\gamma \rangle_0 - \frac{r}{1-(1-r)  \mu^\gamma } \right] (1-r)^t  \mu^{\gamma t} ,
\end{equation}
with $\langle x^\gamma \rangle_0=1$ for $x_0=1$. In contrast with the probability   $p_t^{(m)}$ which, for asymptotically long times, always attains the stationary form given by  Eq.~(\ref{stat0}), the moment $\langle x^\gamma \rangle_t$ converges to a finite value, namely
\begin{equation}
    \langle x^\gamma \rangle_\infty =\frac{r  }{1-(1-r)    \mu^\gamma },
\end{equation}
only if $(1-r)   \mu^\gamma  <1$.  The convergence is exponential in time, within a typical time scale $ |\ln [  (1-r)   \mu^\gamma ]  |^{-1}$.   On the contrary, if $(1-r)   \mu^\gamma \ge 1$, $\langle x^\gamma \rangle_t$   diverges exponentially with time. This convergence or divergence of the moment $ \langle x^\gamma \rangle_t$ depending on whether the order $\gamma$ is respectively lower or larger than the critical value $\gamma_{\rm c} = | \ln(1-r) /\ln \mu|$ is compatible with a probability distribution (for a {\em continuous} variable $x$) decaying as the power law $x^{-1-\gamma_{\rm c}}$. In fact, starting from the stationary distribution of Eq.~(\ref{stat0}), extending the index $m = \ln x/\ln \mu$ to the real positive domain, and changing variables from $m$ to $x$, we get the stationary (``fat-tailed'') power-law  probability distribution
\begin{equation} \label{stat0f}
 f_\infty (x)  = \frac{r}{\ln \mu} x^{-1-|\ln (1-r)/\ln \mu|}.
\end{equation}

\subsection{Dissecting the process: properties of individual realizations}

In certain applications, it is necessary to ``dissect'' each possible realization of the stochastic process, determining the probability of each different way in which the system can evolve. Specifically, for Eq.~(\ref{ev-n}), we first ask what is the probability that, up to a given time $\tau$, the evolution consists of a succession of exactly $k$ bursts of multiplicative growth ($k=1,\dots ,\tau+1$, including one-step bursts between two contiguous reset events). Equivalently, we ask for the probability  that exactly $k-1$ resets occur  somewhere between times $t=1$ and $\tau$, both inclusive. For $k=1$, the probability that no reset has occurred is $(1-r)^\tau$. In turn, each one of the $\tau$ options for one reset ($k=2$) has  probability $(1-r)^{\tau-1} r$. Generally, the probability that the evolution up to time $\tau$ includes $k-1$ resets, i.e. $k$ bursts, is
\begin{equation}
\rho_\tau^{(k)}= \left( \begin{array}{c} \tau \\ k-1 \end{array} \right)
(1-r)^{\tau-k+1} r^{k-1},
\end{equation}
for $0\le k\le \tau$. 

Now, consider the set of realizations where, up to time $\tau$, the evolution   consists of exactly $k$ bursts. How many of all these bursts have a duration of $m+1$ steps ($m=0,1,\dots,\tau$), i.e. at how many times does $x_t$ reach the value $\mu^m$ just before the burst ends (either by a reset event or because time $\tau$ has been reached)? This number can be calculated by considering the so-called {\em compositions} of $\tau$ into $k$ parts,\cite{compos} and turns out to be
\begin{equation}
    K_\tau^{(k,m)} =\left\{ \begin{array}{ll}
   1 & \mbox{for $k=1$ and $m=\tau$}, \\ \\
    k \left( \begin{array}{c} \tau-m-1  \\ k-2 \end{array} \right) & 
    \mbox{for $1<k\le \tau-m+1$ }, \\ \\
    0 & \mbox{otherwise}.
\end{array}
\right.
\end{equation}
Taking into account that, if $x_t$ reaches the value $\mu^m$ during a burst, all the lower values $1,\mu,\mu^2,\dots, \mu^{m-1}$ have also been previously attained, the total number of times that $x_t=\mu^m$ has been attained at any moment $t$ up to time $\tau$, in all the realizations with exactly $k$ bursts,  can be calculated from $K_\tau^{(k,m)}$, yielding
\begin{equation}
    M_\tau^{(k,m)} =\left\{ \begin{array}{ll}
    k \left( \begin{array}{c} \tau-m   \\ k-1 \end{array} \right) & 
    \mbox{for $1\le k \le \tau-m+1$ }, \\ \\
    0 & \mbox{otherwise}.
\end{array}
\right.
\end{equation}
Finally, if the system has evolved up to time $\tau$, the probability that the value of $x_t$ is $\mu^m$ at a randomly chosen time $t\le \tau$ reads
\begin{equation} \label{21}
    \sum_{k=1}^{\tau+1} \rho_\tau^{(k)} \frac{M_t^{(k,m)}}{\sum_{m'=0}^{\tau} M_t^{(k,m')} } = \frac{1+r (\tau-m)}{1+\tau}  (1-r)^m, 
\end{equation}
which, up to the normalization factor $(1+\tau)^{-1}$, corresponds to Eq.~(39) in Ref.~39. Note carefully the difference between  Eqs.~(\ref{16}) and  (\ref{21}). While the former gives the probability that a given value of $x$ is attained after $t$ evolution steps, the latter is the probability constructed by recording the frequency of all the values of $x$ along the whole evolution until time $\tau$. For $t=\tau$, the two expressions coincide in the limit $\tau\to \infty$ only; cf.~Eq.~(\ref{stat0}).

\subsection{Remarks on self-averaging and ergodicity}

It is well known that systems involving stochastic variables with fat-tailed distributions exhibit peculiar statistical features, directly associated with the fact that some of the leading distribution moments ---such as the mean value and/or the variance--- are not finite.\cite{montroll:1982,sornette:1998PRE} Although, for any finite time $t$, the moments for the variable $x_t$ in the stochastic process (\ref{ev-n}) are all finite, as given by Eq.~(\ref{mom-n}), similar features manifest themselves as time elapses and a  fat-tailed asymptotic distribution progressively builds up. 

\begin{figure}[ht]
\centering
\includegraphics[width=.8\columnwidth]{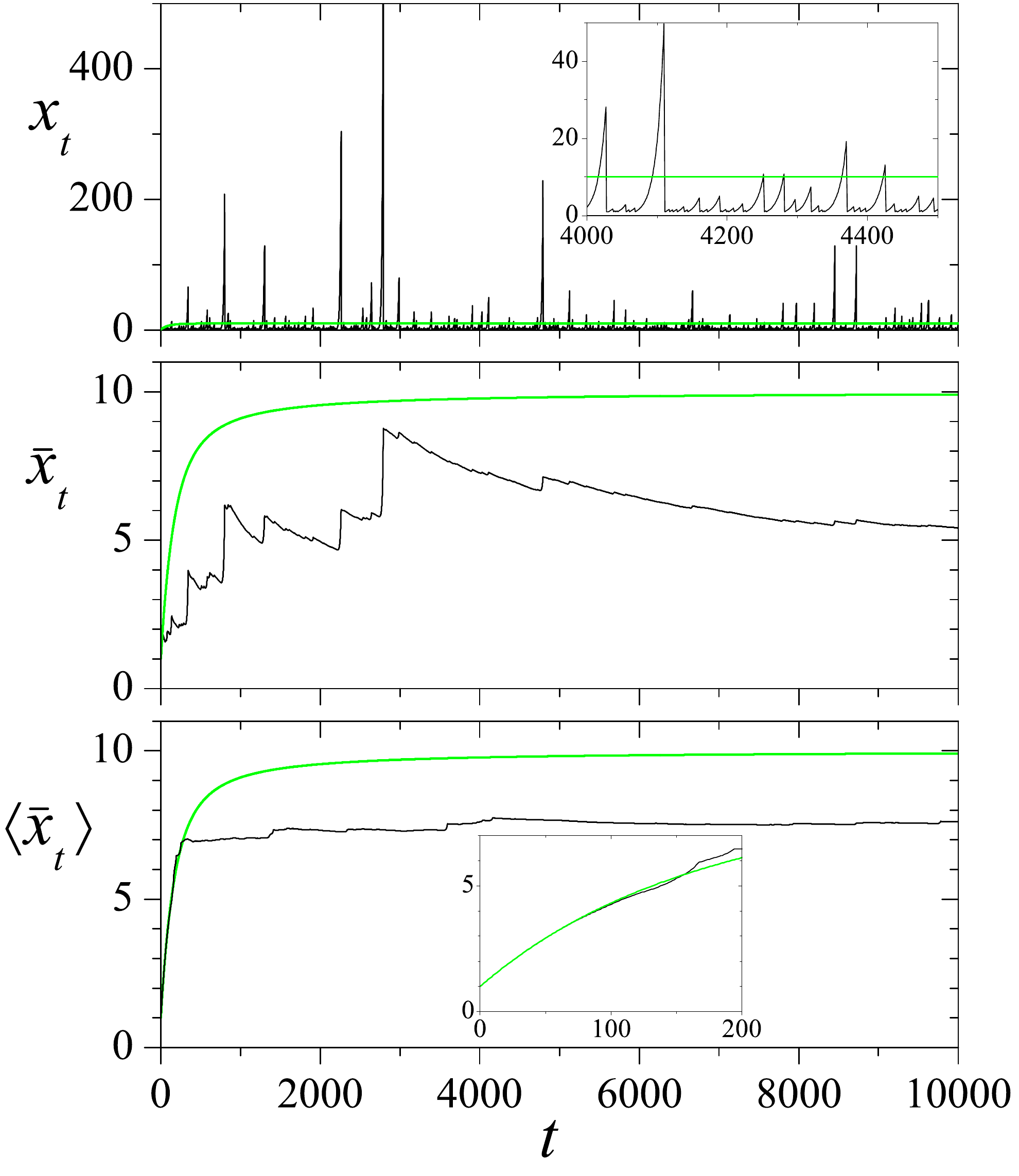}
\caption{Upper panel: $x_t$ as a function of time, along a single realization of the stochastic process (\ref{ev-n}) for $\mu=1.1$ and $r=0.1$ (dark curve). The light-shaded curve stands for  the analytical prediction for $\langle x\rangle_t$, Eq.~(\ref{mom-n}). The inset shows a close-up for intermediate times. Middle panel: Cumulative average of $x_t$ (dark curve; see main text) for the same realization as in the upper panel, and the corresponding analytical prediction, Eq.~(\ref{momp}) (light-shaded curve). Lower panel: As in the middle panel, averaging over $10^6$ realizations of the stochastic process. The inset shows the result for short times.}
\label{fig2}
\end{figure}

Lack of self-averaging ---namely, the failure of a scaled sum of stochastic variables to converge to a well-defined value--- is intuitive for distributions with divergent mean value. This phenomenon has been recognized as a typical feature of random multiplicative processes,\cite{redner} and has been widely discussed for a variety of systems of physical interest,\cite{derrida:1997,derrida:1999} including Markov processes with resets.\cite{meilahn:2015} Self-averaging also fails when the mean value is finite but the variance diverges, a situation relevant to the theory of financial revenues and related problems in economics.\cite{taleb:2007AS,taleb:2008} The fat-tailed asymptotic distribution of process (\ref{ev-n})  has finite mean value $\langle x \rangle$ and divergent variance $\sigma^2 = \langle x^2 \rangle- \langle x \rangle^2$ for $1<\gamma_{\rm c} =  | \ln(1-r) /\ln \mu| <2$.  The upper panel of Fig.~\ref{fig2} shows $x_t$, up to $t=10^4$, along a single realization of the process with $\mu=1.1$ and $r=0.1$, for which $\gamma_{\rm c} \approx 1.1$. The light-shaded curve is the  prediction of Eq.~(\ref{mom-n}) for the mean value (i.e., with $\gamma=1$), which approaches  $\langle x \rangle_\infty =10$ for long times. The large fluctuations of $x_t$ around its expected average are apparent in the main plot and in the inset, which shows a detail for intermediate times. 

The middle panel of Fig.~\ref{fig2} shows the accumulated average of the stochastic variable, defined as $\bar x_t = (t+1)^{-1} \sum_{\tau=0}^t x_\tau$, for the same realization as in the upper panel. The light-shaded curve stands for the corresponding analytical value,  
\begin{eqnarray} 
{ \langle \bar x \rangle}_t  = && \frac{r }{1-(1-r) \mu} \nonumber \\ && + \left[1  - \frac{r}{1-(1-r)\mu} \right] \frac{1-(1-r)^{t+1} \mu^{t+1}}{(t+1)[1-(1-r) \mu]},    \label{momp}
\end{eqnarray}
derived from Eq.~(\ref{mom-n}). Comparison of the two curves puts in evidence, in particular, the role of large values of $x_t$ in building up the cumulative average. Their contribution, in fact, punctuates with abrupt upward jumps the otherwise decaying evolution of $\bar x_t$. It is clear, however, that ---although the typical convergence  time for the mean value, $|\ln [(1-r)\mu]|^{-1} \approx 100$, has long passed---  $\bar x_t$ is nowhere close to $\langle x \rangle_\infty$. This non-self-averaging effect remains even if an additional average is performed over many realizations of the stochastic process ---as implicit  in Eqs.~(\ref{mom-n}) and (\ref{momp}). In the lower panel of Fig.~\ref{fig2} we plot $\bar x_t$ averaged over $10^6$ realizations. While, as shown in the inset, the coincidence between numerical and analytical  results is very good for small $t$, a sustained sizable discrepancy ---which decreases only very slowly as time elapses--- persists for long times.

Somehow disappointingly, there is no formal theory to describe the behavior of the mean value of a stochastic variable ---calculated as the average of successive random draws from a prescribed distribution--- when its variance is infinite. Heuristic approaches based on the estimation of bootstrap distributions or the use of surrogate tail-trimmed variables, however, have been advanced in applications to finance.\cite{boot} A different viewpoint is provided by the extreme-value theory, which focuses on the statistics of the rare events where the variable exhibits severe deviations from the mean value.\cite{redner,rare} In our case, they correspond to the high bursts in the upper panel of Fig.~\ref{fig2}, which cause the sharp steps in the middle panel. 

For the stochastic process (\ref{ev-n}) we can ask what is the mean waiting time $w_X$ until $x_t$ reaches, for the first time, a given (large) value $X =\mu^M$. This event happens at the $M$-th step of a sufficiently long burst, assuming that all the preceding bursts have been shorter than $M$ steps. The corresponding waiting time is the total duration of all the preceding bursts plus $M$ additional steps. Its mean value turns out to be
\begin{equation}  
w_X  =   \frac{1}{r} \left[ \frac{1}{(1-r)^M}-1 \right]   \approx \frac{1}{r} X^{\gamma_{\rm c}},
\end{equation}
where the approximation holds for large $X$, i.e. for large $M$; cf.~Eq.~(18) in Ref.~39.

On the other hand, the mean number $d_X$ of random draws from the stationary distribution (\ref{stat0}) until the value $X=\mu^M$ is obtained, assuming that all the preceding draws have produced lower values, can be immediately found:
\begin{equation}
d_X =  \frac{r}{(1-r)^M} = r X^{\gamma_{\rm c}}.
\end{equation} 
Note that the waiting time $w_X$ is larger than $d_X$ by a factor $r^{-2}$ ($=100$ in the realizations of Fig.~\ref{fig2}). This difference can be viewed as a form of ``transient non-ergodicity'' in our Markov chain. Along the stochastic process, in fact, observation of $x_t=X$ requires waiting the total duration of a succession of shorter bursts where $x_t$ adopts one or more times each value lower than $X$. Instead, when drawing the stochastic variable directly from the distribution, the occurrence of $X$ is independent of the preceding draws. Evaluation of the distribution from the time evolution of $x_t$, consequently, can be much slower than from random realizations of the stochastic variable.     

\section{\label{sec3}The non-uniform case} 
 
Although the stochastic process (\ref{ev-n}) already captures the key mechanisms that lead to the generation of a power-law distribution for the variable $x_t$, it is useful to introduce a couple of generalizations that relax some assumptions implicit in the formulation of the above multiplicative process but, at the same time, preserve most of its analytical tractability. In particular, we now admit that the multiplicative coefficient $\mu$ and the value of $x_t$ after each reset event vary randomly with time. Namely, we consider the Markov chain\cite{manrubia:1999}
\begin{equation} \label{ev-nd}
    x_{t+1}=\left\{
    \begin{array}{ll}
    \mu_t x_t    & \mbox{with probability $1-r$},  \\
       s_t  &   \mbox{with probability $r$},
    \end{array}\right.
\end{equation}
where, at each time step, $\mu_t>0$ and $s_t>0$ are drawn from distributions  $P(\mu)$ and $F(s)$, respectively. In this variant, in contrast with Eq.~(\ref{ev-n}), $x_t$ generally adopts continuous positive values, and its probability is described by a distribution $f_t(x)$. The initial value $x_0$ is chosen from a prescribed distribution $f_0(x)$. The previous case is reobtained taking delta-like profiles for $P(\mu)$, $F(s)$, and $f_0(x)$.

In order to avoid hindering our presentation with the discussion of special pathological situations, we assume that the moments of the distributions $P(\mu)$ and $F(s)$,
\begin{equation} \label{mom}
\langle \mu^\gamma \rangle  =\int_0^\infty \mu^\gamma P (\mu) d\mu, \ \ \ \ \ \langle s^\gamma \rangle  =\int_0^\infty s^\gamma F (s) ds,
\end{equation}
are finite for any order $\gamma  \in (-\infty , \infty)$. This is the case if $P(\mu)$ and $F(s)$ drop rapidly enough to zero for both $\mu, s\to 0$ and $\mu, s \to \infty$. In typical applications, in particular,  the reset values $s_t$ are restricted to some finite domain, so that the support of the distribution $F(s)$ is a bounded interval, and  the finiteness of $\langle s^\gamma \rangle$ is guaranteed for all $\gamma$. The results that we obtain below, however, do not always require that $\langle \mu^\gamma \rangle $ and  $\langle s^\gamma \rangle$ are finite.

In order to obtain the solution of the stochastic process (\ref{ev-nd}), it is convenient to consider the evolution of the logarithmic variable $y_t = \ln x_t$:
\begin{equation} \label{ev-log}
    y_{t+1}=\left\{
    \begin{array}{ll}
    y_t+  \nu_y    & \mbox{with probability $1-r$},  \\
       u_t  &   \mbox{with probability $r$},
    \end{array}\right.
\end{equation}
where $\nu_t =\ln \mu_t$ and $u_t = \ln s_t$. The stochastic variables $\nu_t$ and $u_t$ are respectively drawn from distributions $Q(\nu )$ and $G(u )$, with $Q(\nu) d\nu =P(\mu) d\mu$ and $G(u) du =F(s) ds$. In other words, $Q(\nu) =P(\exp \nu) \exp \nu$ and $G(u) = F(\exp u) \exp u$. The same transformation yields for the initial distribution $g_0(y)$ in terms of $f_0(x)$. The Markov chain for $y_t$, Eq.~(\ref{ev-log}), is an additive (generally, drift plus diffusion) stochastic process with resets.

The  Chapman-Kolmogorov equation for the probability distribution of $x_t$, 
$f_t(x)$, is
\begin{equation}  \label{ev-t}
    f_{t+1} (x)= (1-r) \int_0^\infty  P(\mu) \mu^{-1} f_t  \left( \mu^{-1} x \right) d\mu + r F(x) . 
\end{equation}
Correspondingly, the probability distribution for $y_t$, $g_t (y)$, satisfies
\begin{equation} \label{ev-g}
   g_{t+1} (y)= (1-r) \int_{-\infty}^\infty  Q(\nu) g_t (y-\nu) d\nu + r G(y) . 
\end{equation}
In the Fourier representation, which we define as $\hat g_t (\eta) = \int_{-\infty}^\infty g_t (y) \exp (-2\pi i \eta y ) d y$ and analogous expressions for all the other functions of $y$, the solution to Eq.~(\ref{ev-g}) reads
\begin{eqnarray} 
\hat g_t(\eta)  && =  \frac{r\hat G(\eta)}{1-(1-r)\hat Q(\eta)} \nonumber\\
&& +\left[ \hat g_0 (\eta)-\frac{r\hat G(\eta)}{1-(1-r)\hat Q(\eta)} \right]
(1-r)^t\hat Q(\eta)^t.\label{hat-g}
\end{eqnarray}
Since, up to a multiplicative constant in its variable, $\hat Q (\eta)$ is the characteristic function of the variable $\nu$, it satisfies $|\hat Q (\eta)|  \leq 1$ for all $\eta$. Therefore,  $\hat g_t(\eta)$ converges to the stationary solution
\begin{equation}  \label{stat1}
\hat g_\infty (\eta) =\frac{r\hat G(\eta)}{1-(1-r) \hat Q(\eta)} 
\end{equation}
for $t\to \infty$. 
  
Except for the term including the initial condition, the solution $ \hat g_t (\eta)$ in Eq.~(\ref{hat-g}) is directly proportional to $\hat G (\eta)$, the Fourier transform of the distribution of the stochastic variable immediately after each reset event ---see second line of Eq.~(\ref{ev-log}). The same happens with the stationary distribution $\hat g_\infty (\eta)$ in Eq.~(\ref{stat1}). This proportionality reveals that $g_t(y)$ and, correspondingly, $f_t(x)$ are given ---up to a term involving their initial values--- by a linear superposition of contributions coming from each possible value of the variable after resets, which act as mutually independent ``starts'' for the ensuing multiplicative process. Upon Fourier antitransforming, in fact, both $g_t(y)$ and $f_t(x)$ would be respectively given by convolutions of $G(u)$ and $F(s)$ with distributions representing the contribution of each $u$ and $s$. The effect of having admitted that after resets the variable can adopt different values is, therefore, rather straightforward. 

By Fourier antitransforming Eq.~(\ref{hat-g}), it is in principle possible to find the solution $g_t(y)$ to equation (\ref{ev-g}). In turn, using the identity $g_t(y) dy = f_t(x) dx$, we would obtain the solution  to   equation (\ref{ev-t}), $f_t (x) =x^{-1} g_t(\ln x) $. For generic forms of $Q(\nu)$ and $G(u)$, however, this calculation can seldom be explicitly performed. On the other hand, it is straightforward to exactly find the moments  of the distribution $f_t (x)$, 
\begin{equation}
    \langle x^\gamma \rangle_t =\int_0^\infty x^\gamma f_t (x) dx,
\end{equation}
for any order $\gamma$, by noting that $\langle x^\gamma \rangle_t= \hat g_t (i\gamma/2\pi)$. Evaluating  Eq.~(\ref{hat-g}) at $\eta=i\gamma/2\pi$, in fact, we get
\begin{eqnarray}
  \langle x^\gamma \rangle_t && =\frac{r\langle s^\gamma\rangle}{1-(1-r)\langle \mu^\gamma \rangle}
   \nonumber\\ &&+ \left[\langle x^\gamma \rangle_0 - \frac{r\langle s^\gamma \rangle}{1-(1-r)\langle \mu^\gamma \rangle} \right] (1-r)^t \langle \mu^\gamma \rangle^t, \label{mom-nd}
   \end{eqnarray}
which generalizes Eq.~(\ref{mom-n}). Note that, except for the term involving the initial condition, $\langle x^\gamma \rangle_t$ is proportional to the corresponding moment of the distribution of values after resets, $F(s)$.

\begin{figure}[ht]
\centering
\includegraphics[width=\columnwidth]{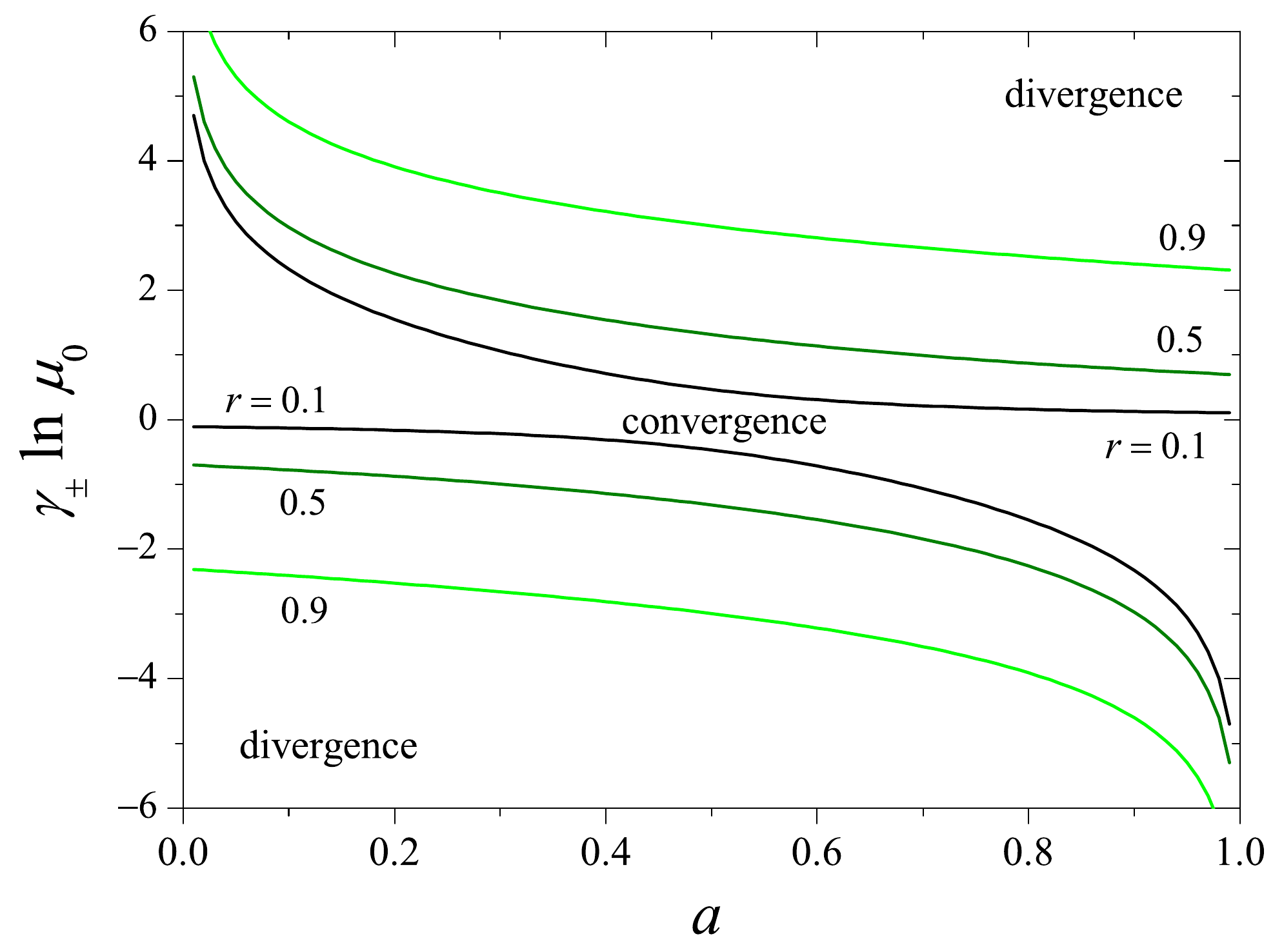}
\caption{The ends $\gamma_\pm$ of the interval of convergence of the moment $\langle x^\gamma \rangle_t$ for the  multiplicative process with resets (\ref{ev-nd}), with $\mu$ drawn from the distribution $P(\mu) = a \delta(\mu-\mu_0)+(1-a) \delta(\mu-\mu_0^{-1})$ and $\mu_0>1$. The values of $\gamma_\pm$ are scaled by $\ln \mu_0$, and plotted as functions of $a$  for three values of the reset probability $r$.}
\label{fig3}
\end{figure}

Much as for the stochastic process (\ref{ev-n}), while the distribution for   $x_t$ always attains a well-defined asymptotic form, given by Eq.~(\ref{stat1}) in the Fourier representation for the logarithmic variable $y_t$, the moment $\langle x^\gamma \rangle_t$ converges to a finite value, 
\begin{equation}
    \langle x^\gamma \rangle_\infty =\frac{r   \langle s^\gamma \rangle}{1-(1-r)   \langle \mu^\gamma \rangle},
\end{equation}
only if $(1-r) \langle \mu^\gamma \rangle <1$. It is interesting to note that ---while, in the uniform case considered in Section \ref{sec2}, convergence or divergence of     $ \langle x^\gamma \rangle_t$ depends on whether $\gamma$ is respectively lower or larger than a critical value $\gamma_{\rm c}$--- in the present case it can happen that the moment converges for $\gamma$ inside a finite interval ($\gamma_-,\gamma_+$), with $\gamma_-<0<\gamma_+$, and diverges elsewhere. This requires, in particular, that the distribution $P(\mu)$ allows for values of $\mu$ at both sides of $\mu=1$. Take, for instance, the two-delta distribution $P(\mu) = a \delta(\mu-\mu_0)+(1-a) \delta(\mu-\mu_0^{-1})$ with $0<a<1$ and $\mu_0\neq 1$. The interval of convergence for the order $\gamma$ is limited by
\begin{equation}
\gamma_\pm = \frac{1}{\ln \mu_0}  \ln \frac{1\pm \sqrt{1-4a(1-a)(1-r)^2}}{2a(1-r)} ,  
\end{equation}
where, without generality loss, we have assumed $\mu_0>1$. Figure \ref{fig3} shows the interval ends $\gamma_\pm$ as  functions of $a$ for three values of $r$.

The long-time convergence or divergence of $ \langle x^\gamma \rangle_t$ for $\gamma $ respectively inside and outside the interval$(\gamma_-,\gamma_+)$ is compatible with a distribution $f_\infty (x)$ which behaves as $x^{-1-\gamma_+}$ for $x\to \infty$ (in agreement with the results of Section \ref{sec2}), and as $x^{-1-\gamma_-}$ for $x\to 0$. Finding its complete form, however, would require to antitransform Eq.~(\ref{stat1}).

\section{\label{sec4}State-dependent multiplication and reset frequency with continuous time}

Coming back to the uniform case considered in Section \ref{sec2},  a continuous-time description can be straightforwardly introduced by first assigning a duration $\Delta t$ to each evolution step. Writing the multiplicative coefficient in the first line of Eq.~(\ref{ev-n}) as $\mu = 1+\lambda \Delta t$ and taking the limit $\Delta t\to 0$, the purely multiplicative part of the stochastic process for $x(t)$ becomes a linear equation, $\dot x = \lambda x$, whose solution grows or decays exponentially, depending on the sign of $\lambda$.  This exponential evolution is punctuated by the reset events, which now occur with frequency (probability per unit time) $q$, related to the reset probability $r$ of  Eq.~(\ref{ev-n}) through $r=q \Delta t$.  As in Section \ref{sec2}, we assume here that the stochastic variable is reset to $x=1$ after each event.

The Chapman-Kolmogorov equation for the probability distribution $f(x,t)$ reads now
\begin{equation} \label{tr1}
\partial_t f + \lambda \partial_x ( x f) = -q f(x,t) + q\delta (x-1) \int_0^\infty f(x',t) dx'. 
\end{equation}
Its interpretation as a continuity equation is transparent. The left-hand side describes probability drift, driven by multiplication, towards larger or smaller values of $x$, depending on $\lambda$ being positive or negative, respectively. The loss and gain terms in the right-hand side, meanwhile, stand for probability sinks and sources associated with reset events. It can be readily seen that Eq.~(\ref{tr1}) preserves the distribution norm $\int_0^\infty f(x,t) dx=1$ at all times, so that the integral in the last term is, in reality, a constant  (see, however, the generalization in Eq. (\ref{tr2}) below). Moreover, if $\lambda >0$ and the support of the initial condition $f_0(x)$ is included in the interval $[1,\infty)$ or, conversely, if $\lambda < 0$ and the support of $f_0(x)$ is included in  $(0,1]$, the distribution $f(x,t)$ will only adopt non-zero values inside either interval at all times. To simplify the presentation, and in agreement with our discussion in Section \ref{sec2}, we assume that $\lambda>0$ and restrict the analysis to the interval  $[1,\infty)$.

\begin{figure}[ht]
\centering
\includegraphics[width=\columnwidth]{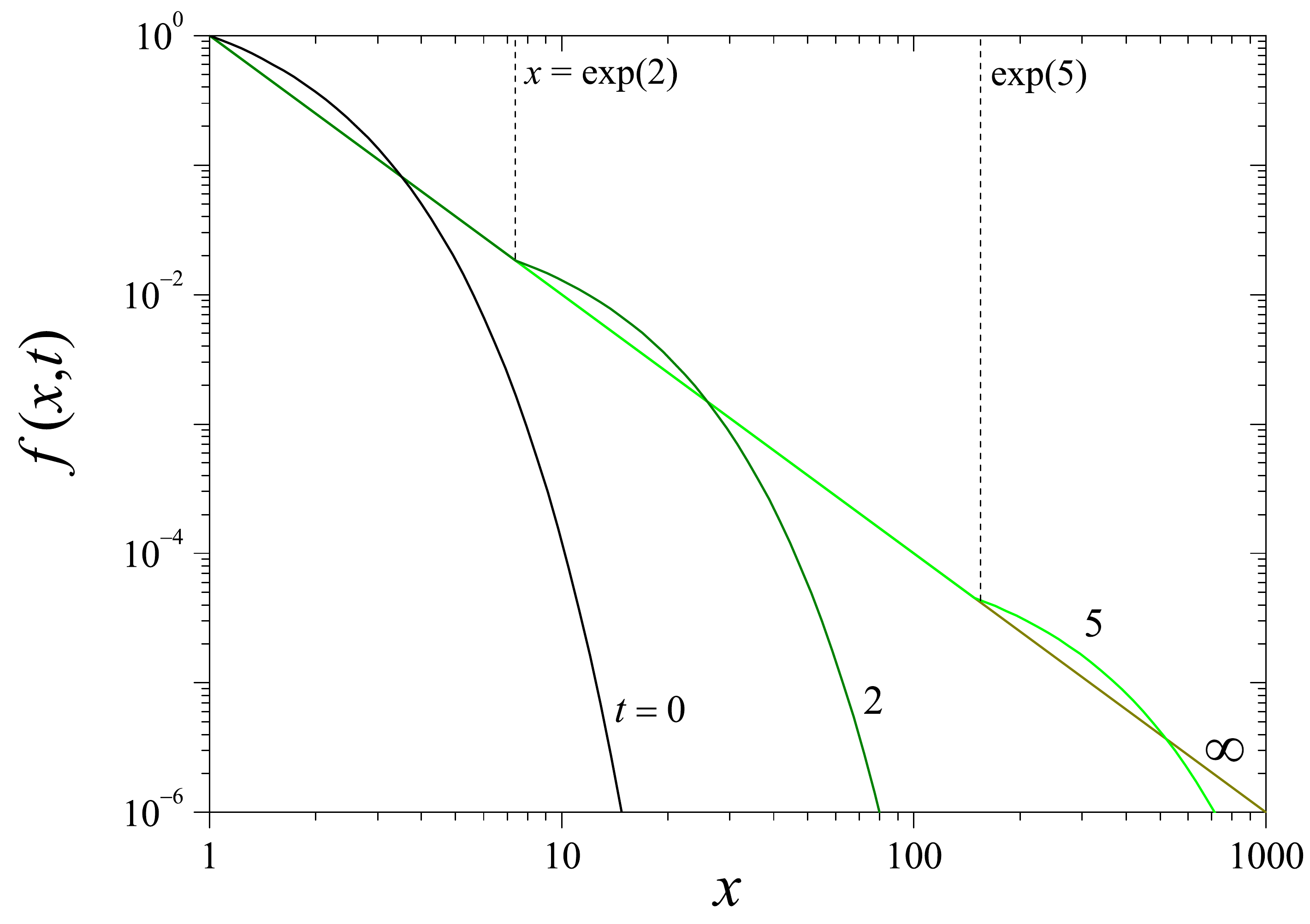}
\caption{The probability distribution $f(x,t)$, Eq.~(\ref{fcont}), as a function of $x$ for $t=0$, $2$, $5$ and $\infty$, starting from an initial condition $f_0(x)= \exp (1-x)$, with $\lambda=q=1$. Dashed vertical lines stand at the boundary $x=\exp (\lambda t)$, which separates the contributions of the initial condition and the asymptotic long-time distribution, $f_\infty (x)=x^{-2}$.}
\label{fig4}
\end{figure}

Under these conditions, Eq.~(\ref{tr1}) can be fully solved by treating the delta-like gain term in the right-hand side as a boundary condition at $x=1$. In fact, assuming that the solution jumps from $f(x,t)=0$ for $x<1$ to a finite value $f(1^+,t)$ just above the boundary, the delta-like term must be balanced by the $x$-derivative in the left-hand side, so that  $f(1^+,t)= q/\lambda$. Taking this condition into account, the solution is
\begin{eqnarray}
   f(x,t) = && \exp [-(\lambda+q)t] f_0[x \exp(-\lambda t)] \Theta [x\exp(-\lambda t)-1] \nonumber \\ && +\frac{q}{\lambda}
   x^{-1-q/\lambda} \Theta [1-x\exp(-\lambda t)] , \label{fcont}
\end{eqnarray}
where $\Theta (x)$ is Heaviside's step function. The distribution $f(x,t)$ is neatly divided into two contributions, corresponding to the two terms in the right-hand side of Eq.~(\ref{fcont}). For $x>\exp (\lambda t)$, in the first term, we have the contribution of the initial condition, which shifts towards increasingly larger values of $x$ and, at the same time, is exponentially  damped and stretched. For $x<\exp (\lambda t)$, in the second term, the asymptotic power-law distribution is established. For $t\to \infty$, this second contribution  spans the whole domain of the variable $x$, yielding
\begin{equation} \label{stat2}
f_\infty (x) =  \frac{q}{\lambda}  x^{-1-q/\lambda},
\end{equation}
to be compared with Eq.~(\ref{stat0f}). As an illustration, Fig.~\ref{fig4} shows, in log-log scale, four snapshots of $f(x,t)$ as a function of $x$, evolving from an initial condition $f_0(x) = \exp (1-x)$, with $x\in [1,\infty)$ and $\lambda=q=1$.  The exponential cutoff originating in $f_0(x)$ shifts to the right as time elapses, while the asymptotic distribution $f_\infty (x)=x^{-2}$ builds up from the left.  
  
Equation (\ref{tr1}) can be immediately generalized to the case where both the coefficient $\lambda$ and the reset frequency $q$ depend on the current value of the stochastic variable $x$. This extension reads
\begin{equation} \label{tr2}
\partial_t f +  \partial_x [ \lambda (x) f] = -q(x) f(x,t) + \delta (x-1) \Phi (t) ,
\end{equation} 
with 
\begin{equation} \label{Phi}
\Phi (t) = \int_0^\infty q(x') f(x',t) dx'. 
\end{equation}
Note that the factor $x$ in the $x$-derivative of Eq.~(\ref{tr1}) has been absorbed by $\lambda (x)$. As in the homogeneous case above, we focus on the problem restricted to the interval $x\in [1,\infty)$, with $\lambda (x)>0$ for all $x$. Treating the last term in the right-hand side of Eq.~(\ref{tr2}) as a boundary condition at $x=1$, we find  
\begin{equation} \label{cond}
f(1^+,t) = \frac{\Phi(t)}{\lambda (1)}.
\end{equation}
In practice, this condition requires to resort to a self-consistent calculation, where $f(x,t)$ is first solved for arbitrary $\Phi(t)$, in such a way that Eq.~(\ref{cond}) is satisfied,  and then $\Phi(t)$ is found from Eq.~(\ref{Phi}). Whether this calculation can be explicitly performed depends on the functional form of $\lambda (x)$ and $q(x)$.

On the other hand, assuming that $f(x,t)$ tends to a well-defined limit for long times, the asymptotic distribution can be readily written as
\begin{equation} \label{statf}
f_\infty (x) = \frac{\Phi_\infty}{\lambda(x)} \exp \left[-\int_1^x \frac{q(x')}{\lambda (x')} dx' \right]    ,
\end{equation}
with $\Phi_\infty$ the asymptotic value of $\Phi(t)$ for $t\to \infty$. Equation (\ref{statf}) makes it clear that $\Phi_\infty$ is essentially fixed by the normalization of $f_\infty (x)$. In fact, it can be easily verified that Eqs.~(\ref{Phi}) to (\ref{statf}) are mutually consistent if the distribution is normalized to unity. Such normalization, however, requires that the integral in the exponential of Eq.~(\ref{statf}) diverges as $x\to \infty$, namely,
\begin{equation} \label{condinf}
    \int_1^\infty \frac{q(x')}{\lambda (x')} dx' = \infty.
\end{equation}
This condition is verified when the reset frequency increases and/or the multiplicative coefficient decreases sufficiently fast as $x$ grows. It expresses the fact that there is no probability ``leaking'' towards large values of $x$ due to excessively weakened resets and/or strengthened multiplication.  

\begin{figure}[ht]
\centering
\includegraphics[width=\columnwidth]{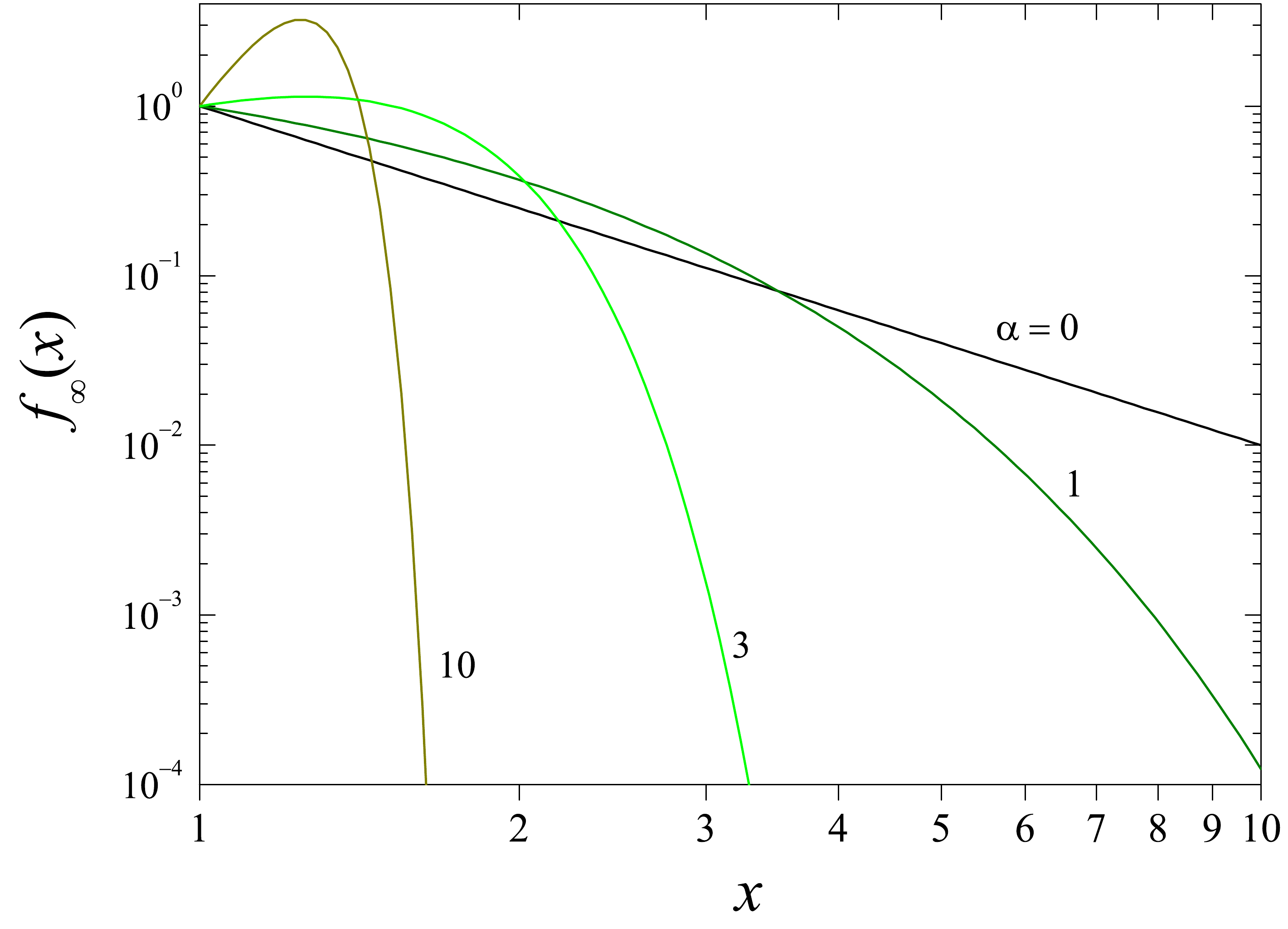}
\caption{The probability distribution of Eq.~(\ref{Wei}) for $\lambda_0=q=1$ and four values of the exponent $\alpha$. For $\alpha=0$, it reduces to the power-law distribution $f_\infty (x) =x^{-2}$.}
\label{fig5}
\end{figure}

From Eq.~(\ref{statf}), moreover, it is apparent that admitting state-dependent multiplication and reset frequency can significantly widen the class of stationary distributions covered by the   model, well beyond power-law decaying functions. As an illustration,  consider the case of constant reset frequency $q$, with a multiplicative coefficient  with algebraic dependence on $x$, namely, $\lambda (x)= \lambda_0 x^{1-\alpha}$. Condition (\ref{condinf}) is fulfilled if $\alpha\ge 0$. For $\alpha>0$, the resulting distribution is
\begin{equation} \label{Wei}
f_\infty (x) = \frac{q  }{\lambda_0 x^{1-\alpha}} \exp\left[- \frac{q(x^\alpha-1)}{\lambda_0 \alpha}\right],
\end{equation}
which reduces to Eq.~(\ref{stat2}) for $\alpha\to 0$. This form of $f_\infty (x)$, shown in Fig.~\ref{fig5} for $\lambda_0=q=1$ and some values of the exponent $\alpha$, is closely related to the Weibull and the stretched exponential distribution, which play a key role in the quantitative description of several socioeconomic systems. The present model, thus, provides a unified mechanism for the occurrence of a variety of distributions relevant to this kind of problems. 

\section{\label{sec5}Discussion and conclusion}

Stochastic multiplicative processes (SMPs) with reset events, in their several variants, are characterized by fat-tailed, often power-law, distributions of the relevant variable. These processes show punctuated behaviour, where periods of growth are terminated by sudden shifts to a previous state. High-order moments of the probability distribution diverge in most variants of the general model. These features are characteristic of, among others, financial processes, a context where higher-order moments are relevant. Kurtosis, for instance, is typically used as a measure of financial risk: divergence of the fourth moment (even with finite variance and skewness) yields unreliable predictions if evaluated using finite data sets.\cite{taleb:2007AS,taleb:2008}  An additional difficulty in the numerical estimation of moments ---and, therefore, of risk--- is due to the very slow convergence of estimated values to the exact values of the process, even if the associated moments are finite. This ``slow law of large numbers'' is caused by the large weight of rare events (black swans), which take a lot of data to show up, and prevent a proper estimation of the moments of such processes through the moments of a sample.~\cite{taleb:2008}

The interpretation of SMPs with resets in the context of finance is quite straightforward. Regardless of whether the model is implemented in discrete or continuous time, the relevant variable, either $x_t$ or $x(t)$, can be understood as the gain accumulated at a given time $t$, while its dynamics are controlled by two parameters, one characterizing the momentary gain ($\mu$ or $\lambda$), and another one quantifying the risk of the investment ($r$ or $q$). The uniform model analyzed in Section \ref{sec2} would therefore describe an ensemble of gamblers devoid of strategy and democratically dependent on luck. Still, they experience dissimilar fortunes and could be, {\it a posteriori}, evaluated as poor, mediocre, or excellent investors. It has been suggested that the search for causality in lucky realizations of random processes, together with the elimination of unlucky series from samples, might lead to highly misleading interpretations of the actual causal mechanisms of data, a phenomenon of particular relevance in finance known as ``survivorship bias.''~\cite{taleb:2005} 

The simple model with constant risk and momentary gain clarifies other aspects of the process as well. The exponent of the power law depends on both parameters: the higher the frequency and magnitude of resets, the smaller the exponent (in absolute value). That is, black swans will have larger impact if the momentary gain increases, and also if the risk is high, as it could have been guessed. Furthermore, the estimation of the waiting time to observe an event of size $X$ shows that the typical, additional waiting time until an event of, say, twice that size occurs is $2^{\gamma_c}$ time steps longer. If  $\gamma_c$ is unknown, or if its estimation is affected by large errors, predictions of the expected time until the next black swan and of its magnitude can be highly inaccurate, at best. 

The non-uniform model in Section \ref{sec3} extends those results by considering a broad class of probability distributions for momentary gains and reset values, while the risk is kept fixed. Momentary gain is now time-dependent and can be larger or smaller than one (in the latter case, it becomes a momentary loss), this generalization leading to two-sided power laws and to moments converging within a finite interval of $\gamma$ values. The qualitative properties of this model are fully comparable to those of the uniform case, though additional situations can be now embraced. The accumulated gain can increase or decrease in random amounts at each time step, and the distribution of reset values $F(s)$ could take into account agents with different strategies regarding the fraction of their gains put at stake. Prudent investors might save part of their previous gains to avoid too severe drops. The details of this strategy directly affect the moments of the accumulated gain distribution, though the power-law function persists.  

Finally, in Section \ref{sec4}, we have studied a  variant where the momentary gain and risk can depend on the accumulated gain. In this case, the model has been formulated in continuous time for the sake of analytical tractability, but can be interpreted along the same lines as discrete-time models. This last case incorporates the possibility that agents, perhaps content with the gains accumulated so far, develop cautious strategies by investing in products with lower momentary gain but which are less risk-prone. Interestingly, this non-greedy strategy can transform the power-law distribution into different fat-tailed distributions, where the effects of black swans might be strongly suppressed. 

A relevant situation that we have not explicitly explored in this work is the plausible relationship between risk and gain. In practice, it comes to reason that financial gain cannot be maximized while simultaneously minimizing risk. For example, if the momentary gain in the uniform model becomes a linear function of risk, $\mu = 1+\kappa r$ ($\kappa>0$), then $\gamma_c \approx 1/\kappa$ for small risk. Under the constraint of mutual dependency, strategies decrease the number of their degrees of freedom, and similar restrictions would hold if the relationship between gain and risk affects their statistical distribution (note that, implicitly, Eq.~(\ref{condinf}) establishes a form of weak constraint linking the two  parameters). Still, the implementation of realistic constraints in scenarios where, furthermore, those distributions depend on the accumulated gain appears as an interesting avenue to explore. In a different context, it has been shown that a reset probability that depends on time can improve the efficiency of search processes.~\cite{kusmierz:2019} Another extension with potential applicability is the consideration of trends in, e.g., the minimum reset value. A variety of situations might be subject to such trends, as Fig.~\ref{fig:PriceIndex} illustrates for food price indices. The trend might be positive (inflation), negative (deflation), or be itself subject to large variations, as it happens with hyperinflation followed by currency devaluation. 

Finance offer multiple situations that can potentially be modelled through SMPs with resets and, occasionally, additional mechanisms tailored to specific  scenarios. Possible applications are however broader and extend beyond the many cases already described in this work.~\cite{beare:2019} In population dynamics, also dominated by multiplicative (demographic) growth processes, the fast, local extinction of a population is often followed by ``reinjection'' in the form of a small number of migrating individuals. This situation could describe parasitic infection bursts in metapopulations~\cite{buendia:2018} and explain the persistence of populations that would otherwise become extinct.~\cite{adler:1993} Resets could be also rephrased as any process that finishes the multiplicative growth, since the properties described do not depend on whether the reset is repeatedly applied to realizations that have a continuity in time or to many different realizations that are independently  ``born,'' and  then terminated at the time of resetting. An example can be found in models for cascade fracture with stopping events.~\cite{yamamoto:2012} Applications of SMPs to such areas as material sciences and population dynamics would require their generalization to spatially extended systems, considering coupled stochastic processes occurring at neighbor sites.

Given the ubiquity of multiplicative processes possibly amended by a variety of mechanisms ---resets, among many others--- the profusion of power-law (or fat-tailed) distributed quantities found in natural and social sciences no longer comes as a surprising fact. Such distributions entail non-trivial dynamical properties, as those characterized in this work, that severely limit our ability to predict future outcomes of the process. Despite the advances of over a century of research on this topic, further applications and deeper analytical approaches are yet to come to improve our understanding of the mechanisms generating fat-tailed distributions and our control of black-swan-like events. 

\section*{Acknowledgements}

S.~M.~is supported by grant FIS2017-89773-P (MINECO/FEDER, E.~U.). 

\bibliography{zanettemanrubia} 

\providecommand{\noopsort}[1]{}\providecommand{\singleletter}[1]{#1}%
\begin{thebibliography}{50}%
\makeatletter
\providecommand \@ifxundefined [1]{%
 \@ifx{#1\undefined}
}%
\providecommand \@ifnum [1]{%
 \ifnum #1\expandafter \@firstoftwo
 \else \expandafter \@secondoftwo
 \fi
}%
\providecommand \@ifx [1]{%
 \ifx #1\expandafter \@firstoftwo
 \else \expandafter \@secondoftwo
 \fi
}%
\providecommand \natexlab [1]{#1}%
\providecommand \enquote  [1]{``#1''}%
\providecommand \bibnamefont  [1]{#1}%
\providecommand \bibfnamefont [1]{#1}%
\providecommand \citenamefont [1]{#1}%
\providecommand \href@noop [0]{\@secondoftwo}%
\providecommand \href [0]{\begingroup \@sanitize@url \@href}%
\providecommand \@href[1]{\@@startlink{#1}\@@href}%
\providecommand \@@href[1]{\endgroup#1\@@endlink}%
\providecommand \@sanitize@url [0]{\catcode `\\12\catcode `\$12\catcode
  `\&12\catcode `\#12\catcode `\^12\catcode `\_12\catcode `\%12\relax}%
\providecommand \@@startlink[1]{}%
\providecommand \@@endlink[0]{}%
\providecommand \url  [0]{\begingroup\@sanitize@url \@url }%
\providecommand \@url [1]{\endgroup\@href {#1}{\urlprefix }}%
\providecommand \urlprefix  [0]{URL }%
\providecommand \Eprint [0]{\href }%
\providecommand \doibase [0]{http://dx.doi.org/}%
\providecommand \selectlanguage [0]{\@gobble}%
\providecommand \bibinfo  [0]{\@secondoftwo}%
\providecommand \bibfield  [0]{\@secondoftwo}%
\providecommand \translation [1]{[#1]}%
\providecommand \BibitemOpen [0]{}%
\providecommand \bibitemStop [0]{}%
\providecommand \bibitemNoStop [0]{.\EOS\space}%
\providecommand \EOS [0]{\spacefactor3000\relax}%
\providecommand \BibitemShut  [1]{\csname bibitem#1\endcsname}%
\let\auto@bib@innerbib\@empty
\bibitem [{\citenamefont {Auerbach}(1913)}]{auerbach:1913}%
  \BibitemOpen
  \bibfield  {author} {\bibinfo {author} {\bibfnamefont {F.}~\bibnamefont
  {Auerbach}},\ }\bibfield  {title} {\enquote {\bibinfo {title} {Das {Gesetz
  der Bev\"olkerungskonzentration}},}\ }in\ \href@noop {} {\emph {\bibinfo
  {booktitle} {Petermanns Geographische Mitteilungen}}},\ Vol.~\bibinfo
  {volume} {59},\ \bibinfo {editor} {edited by\ \bibinfo {editor}
  {\bibfnamefont {P.}~\bibnamefont {Langhans}}}\ (\bibinfo  {publisher} {Justus
  Pertes},\ \bibinfo {address} {Gotha},\ \bibinfo {year} {1913})\ pp.\ \bibinfo
  {pages} {74--76},\ \bibinfo {note} {available online at
  https://archive.org/details/Auerbach1913. Translation of first paragraph by
  S.M.}\BibitemShut {Stop}%
\bibitem [{\citenamefont {Estoup}(1916)}]{estoup:1916}%
  \BibitemOpen
  \bibfield  {author} {\bibinfo {author} {\bibfnamefont {J.~B.}\ \bibnamefont
  {Estoup}},\ }\href@noop {} {\emph {\bibinfo {title} {Gammes
  st\'enographiques}}}\ (\bibinfo  {publisher} {Institut St\'enographique de
  France},\ \bibinfo {address} {Paris},\ \bibinfo {year} {1916})\BibitemShut
  {NoStop}%
\bibitem [{\citenamefont {Yule}(1924)}]{yule:1924}%
  \BibitemOpen
  \bibfield  {author} {\bibinfo {author} {\bibfnamefont {G.~U.}\ \bibnamefont
  {Yule}},\ }\bibfield  {title} {\enquote {\bibinfo {title} {A mathematical
  theory of evolution, based on the conclusions of {Dr. J. C. Willis}},}\
  }\href@noop {} {\bibfield  {journal} {\bibinfo  {journal} {Proc. R. Soc.
  London B}\ }\textbf {\bibinfo {volume} {213}},\ \bibinfo {pages} {21--87}
  (\bibinfo {year} {1924})},\ \bibinfo {note}
  {https://doi.org/10.1098/rstb.1925.0002}\BibitemShut {NoStop}%
\bibitem [{\citenamefont {Zipf}(1949)}]{zipf:1949}%
  \BibitemOpen
  \bibfield  {author} {\bibinfo {author} {\bibfnamefont {G.~K.}\ \bibnamefont
  {Zipf}},\ }\href@noop {} {\emph {\bibinfo {title} {Human Behavior and the
  Principle of Least Effort}}}\ (\bibinfo  {publisher} {Addison-Wesley},\
  \bibinfo {address} {Cambridge, Massachusetts},\ \bibinfo {year}
  {1949})\BibitemShut {NoStop}%
\bibitem [{\citenamefont {Simon}(1955)}]{simon:1955}%
  \BibitemOpen
  \bibfield  {author} {\bibinfo {author} {\bibfnamefont {H.~A.}\ \bibnamefont
  {Simon}},\ }\bibfield  {title} {\enquote {\bibinfo {title} {On a class of
  skew distribution functions},}\ }\href@noop {} {\bibfield  {journal}
  {\bibinfo  {journal} {Biometrika}\ }\textbf {\bibinfo {volume} {42}},\
  \bibinfo {pages} {425--440} (\bibinfo {year} {1955})},\ \bibinfo {note}
  {https://doi.org/10.1093/biomet/42.3-4.425}\BibitemShut {NoStop}%
\bibitem [{\citenamefont {Taleb}(2007{\natexlab{a}})}]{taleb:2007}%
  \BibitemOpen
  \bibfield  {author} {\bibinfo {author} {\bibfnamefont {N.~N.}\ \bibnamefont
  {Taleb}},\ }\href@noop {} {\emph {\bibinfo {title} {The Black Swan: The
  Impact of the Highly Improbable}}}\ (\bibinfo  {publisher} {Random House},\
  \bibinfo {address} {New York},\ \bibinfo {year} {2007})\BibitemShut {NoStop}%
\bibitem [{\citenamefont {Manrubia}\ and\ \citenamefont
  {Zanette}(1999)}]{manrubia:1999}%
  \BibitemOpen
  \bibfield  {author} {\bibinfo {author} {\bibfnamefont {S.~C.}\ \bibnamefont
  {Manrubia}}\ and\ \bibinfo {author} {\bibfnamefont {D.~H.}\ \bibnamefont
  {Zanette}},\ }\bibfield  {title} {\enquote {\bibinfo {title} {Stochastic
  multiplicative processes with reset events},}\ }\href@noop {} {\bibfield
  {journal} {\bibinfo  {journal} {Phys. Rev. E}\ }\textbf {\bibinfo {volume}
  {59}},\ \bibinfo {pages} {4945--4948} (\bibinfo {year} {1999})},\ \bibinfo
  {note} {https://doi.org/10.1103/PhysRevE.59.4945}\BibitemShut {NoStop}%
\bibitem [{\citenamefont {Evans}\ and\ \citenamefont
  {Majumdar}(2011)}]{evans:2011}%
  \BibitemOpen
  \bibfield  {author} {\bibinfo {author} {\bibfnamefont {M.~R.}\ \bibnamefont
  {Evans}}\ and\ \bibinfo {author} {\bibfnamefont {S.~N.}\ \bibnamefont
  {Majumdar}},\ }\bibfield  {title} {\enquote {\bibinfo {title} {Diffusion with
  stochastic resetting},}\ }\href@noop {} {\bibfield  {journal} {\bibinfo
  {journal} {Phys. Rev. Lett.}\ }\textbf {\bibinfo {volume} {106}},\ \bibinfo
  {pages} {160601} (\bibinfo {year} {2011})},\ \bibinfo {note}
  {https://doi.org/10.1103/PhysRevLett.106.160601}\BibitemShut {NoStop}%
\bibitem [{\citenamefont {Montero}\ and\ \citenamefont
  {Villarroel}(2013)}]{montero:2013}%
  \BibitemOpen
  \bibfield  {author} {\bibinfo {author} {\bibfnamefont {M.}~\bibnamefont
  {Montero}}\ and\ \bibinfo {author} {\bibfnamefont {J.}~\bibnamefont
  {Villarroel}},\ }\bibfield  {title} {\enquote {\bibinfo {title} {Monotonic
  continuous-time random walks with drift and stochastic reset events},}\
  }\href@noop {} {\bibfield  {journal} {\bibinfo  {journal} {Phys. Rev. E}\
  }\textbf {\bibinfo {volume} {87}},\ \bibinfo {pages} {012116} (\bibinfo
  {year} {2013})},\ \bibinfo {note}
  {https://doi.org/10.1103/PhysRevE.87.012116}\BibitemShut {NoStop}%
\bibitem [{\citenamefont {Pareto}(1896)}]{pareto:1896}%
  \BibitemOpen
  \bibfield  {author} {\bibinfo {author} {\bibfnamefont {V.}~\bibnamefont
  {Pareto}},\ }\href@noop {} {\emph {\bibinfo {title} {Cours d’\'economie
  politique}}}\ (\bibinfo  {publisher} {Rouge},\ \bibinfo {address}
  {Lausanne},\ \bibinfo {year} {1896})\BibitemShut {NoStop}%
\bibitem [{\citenamefont {Newman}(2004)}]{newman:2004}%
  \BibitemOpen
  \bibfield  {author} {\bibinfo {author} {\bibfnamefont {M.~E.~J.}\
  \bibnamefont {Newman}},\ }\bibfield  {title} {\enquote {\bibinfo {title}
  {Power laws, {P}areto distributions and {Z}ipf's law},}\ }\href@noop {}
  {\bibfield  {journal} {\bibinfo  {journal} {Contemp. Phys.}\ }\textbf
  {\bibinfo {volume} {46}},\ \bibinfo {pages} {323--351} (\bibinfo {year}
  {2004})},\ \bibinfo {note}
  {https://doi.org/10.1080/00107510500052444}\BibitemShut {NoStop}%
\bibitem [{\citenamefont {Mitzenmacher}(2004)}]{MM}%
  \BibitemOpen
  \bibfield  {author} {\bibinfo {author} {\bibfnamefont {M.}~\bibnamefont
  {Mitzenmacher}},\ }\bibfield  {title} {\enquote {\bibinfo {title} {A brief
  history of generative models for power law and lognormal distributions},}\
  }\href@noop {} {\bibfield  {journal} {\bibinfo  {journal} {Internet Math.}\
  }\textbf {\bibinfo {volume} {1}},\ \bibinfo {pages} {226--251} (\bibinfo
  {year} {2004})},\ \bibinfo {note}
  {https://doi.org/10.1080/15427951.2004.10129088}\BibitemShut {NoStop}%
\bibitem [{\citenamefont {Zanette}\ and\ \citenamefont
  {Manrubia}(2007)}]{zanette:2007}%
  \BibitemOpen
  \bibfield  {author} {\bibinfo {author} {\bibfnamefont {D.~H.}\ \bibnamefont
  {Zanette}}\ and\ \bibinfo {author} {\bibfnamefont {S.~C.}\ \bibnamefont
  {Manrubia}},\ }\bibfield  {title} {\enquote {\bibinfo {title} {Multiplicative
  processes in social systems},}\ }in\ \href@noop {} {\emph {\bibinfo
  {booktitle} {Complex Population Dynamics. Nonlinear Modeling in Ecology,
  Epidemiology and Genetics}}},\ \bibinfo {series} {Lecture Notes in Complex
  Systems}, Vol.~\bibinfo {volume} {7},\ \bibinfo {editor} {edited by\ \bibinfo
  {editor} {\bibfnamefont {B.}~\bibnamefont {Blasius}}, \bibinfo {editor}
  {\bibfnamefont {J.}~\bibnamefont {Kurths}}, \ and\ \bibinfo {editor}
  {\bibfnamefont {L.}~\bibnamefont {Stone}}}\ (\bibinfo  {publisher} {World
  Scientific},\ \bibinfo {address} {Singapore},\ \bibinfo {year} {2007})\ pp.\
  \bibinfo {pages} {129--158},\ \bibinfo {note}
  {https://doi.org/10.1142/6600}\BibitemShut {NoStop}%
\bibitem [{\citenamefont {Montroll}\ and\ \citenamefont
  {Shlesinger}(1982)}]{montroll:1982}%
  \BibitemOpen
  \bibfield  {author} {\bibinfo {author} {\bibfnamefont {E.~W.}\ \bibnamefont
  {Montroll}}\ and\ \bibinfo {author} {\bibfnamefont {M.~F.}\ \bibnamefont
  {Shlesinger}},\ }\bibfield  {title} {\enquote {\bibinfo {title} {On 1/f noise
  and other distributions with long tails},}\ }\href@noop {} {\bibfield
  {journal} {\bibinfo  {journal} {Proc. Natl. Acad. Sci. USA}\ }\textbf
  {\bibinfo {volume} {79}},\ \bibinfo {pages} {3380--3383} (\bibinfo {year}
  {1982})},\ \bibinfo {note}
  {https://doi.org/10.1073/pnas.79.10.3380}\BibitemShut {NoStop}%
\bibitem [{\citenamefont {Zanette}(2008)}]{zanette:2018}%
  \BibitemOpen
  \bibfield  {author} {\bibinfo {author} {\bibfnamefont {D.~H.}\ \bibnamefont
  {Zanette}},\ }\bibfield  {title} {\enquote {\bibinfo {title} {Multiplicative
  processes and city sizes},}\ }in\ \href@noop {} {\emph {\bibinfo {booktitle}
  {The Dynamics of Complex Urban Systems. An Interdisciplinary Approach}}},\
  \bibinfo {editor} {edited by\ \bibinfo {editor} {\bibfnamefont
  {S.}~\bibnamefont {Albeverio}}, \bibinfo {editor} {\bibfnamefont
  {D.}~\bibnamefont {Andrey}}, \bibinfo {editor} {\bibfnamefont
  {P.}~\bibnamefont {Giordano}}, \ and\ \bibinfo {editor} {\bibfnamefont
  {A.}~\bibnamefont {vancheri}}}\ (\bibinfo  {publisher} {Physica-Verlag},\
  \bibinfo {address} {Heidelberg},\ \bibinfo {year} {2008})\ pp.\ \bibinfo
  {pages} {457--472},\ \bibinfo {note}
  {https://doi.org/10.1007/978-3-7908-1937-3}\BibitemShut {NoStop}%
\bibitem [{\citenamefont {Fu}\ \emph {et~al.}(2005)\citenamefont {Fu},
  \citenamefont {Pammolli}, \citenamefont {Buldyrev}, \citenamefont
  {Riccaboni}, \citenamefont {Matia}, \citenamefont {Yamasaki},\ and\
  \citenamefont {Stanley}}]{fu:2005}%
  \BibitemOpen
  \bibfield  {author} {\bibinfo {author} {\bibfnamefont {D.}~\bibnamefont
  {Fu}}, \bibinfo {author} {\bibfnamefont {F.}~\bibnamefont {Pammolli}},
  \bibinfo {author} {\bibfnamefont {S.~V.}\ \bibnamefont {Buldyrev}}, \bibinfo
  {author} {\bibfnamefont {M.}~\bibnamefont {Riccaboni}}, \bibinfo {author}
  {\bibfnamefont {K.}~\bibnamefont {Matia}}, \bibinfo {author} {\bibfnamefont
  {K.}~\bibnamefont {Yamasaki}}, \ and\ \bibinfo {author} {\bibfnamefont
  {H.~E.}\ \bibnamefont {Stanley}},\ }\bibfield  {title} {\enquote {\bibinfo
  {title} {The growth of business firms: Theoretical framework and empirical
  evidence},}\ }\href@noop {} {\bibfield  {journal} {\bibinfo  {journal} {Proc.
  Natl. Acad. Sci. USA}\ }\textbf {\bibinfo {volume} {102}},\ \bibinfo {pages}
  {18801--18806} (\bibinfo {year} {2005})},\ \bibinfo {note}
  {https://doi.org/10.1073/pnas.0509543102}\BibitemShut {NoStop}%
\bibitem [{\citenamefont {Zanette}\ and\ \citenamefont
  {Manrubia}(2001)}]{zanette:2001}%
  \BibitemOpen
  \bibfield  {author} {\bibinfo {author} {\bibfnamefont {D.~H.}\ \bibnamefont
  {Zanette}}\ and\ \bibinfo {author} {\bibfnamefont {S.~C.}\ \bibnamefont
  {Manrubia}},\ }\bibfield  {title} {\enquote {\bibinfo {title} {Vertical
  transmission of culture and the distribution of family names},}\ }\href@noop
  {} {\bibfield  {journal} {\bibinfo  {journal} {Physica A}\ }\textbf {\bibinfo
  {volume} {295}},\ \bibinfo {pages} {1--8} (\bibinfo {year} {2001})},\
  \bibinfo {note} {https://doi.org/10.1016/S0378-4371(01)00046-2}\BibitemShut
  {NoStop}%
\bibitem [{\citenamefont {Manrubia}\ and\ \citenamefont
  {Zanette}(2002)}]{manrubia:2002}%
  \BibitemOpen
  \bibfield  {author} {\bibinfo {author} {\bibfnamefont {S.~C.}\ \bibnamefont
  {Manrubia}}\ and\ \bibinfo {author} {\bibfnamefont {D.~H.}\ \bibnamefont
  {Zanette}},\ }\bibfield  {title} {\enquote {\bibinfo {title} {At the boundary
  between biological and cultural evolution: The origin of surname
  distributions},}\ }\href@noop {} {\bibfield  {journal} {\bibinfo  {journal}
  {Journal of Theoretical Biology}\ }\textbf {\bibinfo {volume} {216}},\
  \bibinfo {pages} {461--477} (\bibinfo {year} {2002})},\ \bibinfo {note}
  {https://doi.org/10.1006/jtbi.2002.3002}\BibitemShut {NoStop}%
\bibitem [{\citenamefont {Zanette}\ and\ \citenamefont
  {Montemurro}(2005)}]{zanette:2005}%
  \BibitemOpen
  \bibfield  {author} {\bibinfo {author} {\bibfnamefont {D.~H.}\ \bibnamefont
  {Zanette}}\ and\ \bibinfo {author} {\bibfnamefont {M.}~\bibnamefont
  {Montemurro}},\ }\bibfield  {title} {\enquote {\bibinfo {title} {Dynamics of
  text generation with realistic {Z}ipf's distribution},}\ }\href@noop {}
  {\bibfield  {journal} {\bibinfo  {journal} {J. Quant. Ling.}\ }\textbf
  {\bibinfo {volume} {12}},\ \bibinfo {pages} {29--40} (\bibinfo {year}
  {2005})},\ \bibinfo {note}
  {https://doi.org/10.1080/09296170500055293}\BibitemShut {NoStop}%
\bibitem [{\citenamefont {Zanette}(2006)}]{zanette:2006}%
  \BibitemOpen
  \bibfield  {author} {\bibinfo {author} {\bibfnamefont {D.~H.}\ \bibnamefont
  {Zanette}},\ }\bibfield  {title} {\enquote {\bibinfo {title} {Zipf's law and
  the creation of musical context},}\ }\href@noop {} {\bibfield  {journal}
  {\bibinfo  {journal} {Musicae Scientiae}\ }\textbf {\bibinfo {volume} {10}},\
  \bibinfo {pages} {3--18} (\bibinfo {year} {2006})},\ \bibinfo {note}
  {https://doi.org/10.1177/102986490601000101}\BibitemShut {NoStop}%
\bibitem [{\citenamefont {Takayasu}, \citenamefont {Sato},\ and\ \citenamefont
  {Takayasu}(1997)}]{takayasu:1997}%
  \BibitemOpen
  \bibfield  {author} {\bibinfo {author} {\bibfnamefont {H.}~\bibnamefont
  {Takayasu}}, \bibinfo {author} {\bibfnamefont {A.-H.}\ \bibnamefont {Sato}},
  \ and\ \bibinfo {author} {\bibfnamefont {M.}~\bibnamefont {Takayasu}},\
  }\bibfield  {title} {\enquote {\bibinfo {title} {Stable infinite variance
  fluctuations in randomly amplified {Langevin} systems},}\ }\href@noop {}
  {\bibfield  {journal} {\bibinfo  {journal} {Phys. Rev. Lett.}\ }\textbf
  {\bibinfo {volume} {79}},\ \bibinfo {pages} {966--969} (\bibinfo {year}
  {1997})},\ \bibinfo {note}
  {https://doi.org/10.1103/PhysRevLett.79.966}\BibitemShut {NoStop}%
\bibitem [{\citenamefont {Sornette}(1998{\natexlab{a}})}]{sornette:1998PRE}%
  \BibitemOpen
  \bibfield  {author} {\bibinfo {author} {\bibfnamefont {D.}~\bibnamefont
  {Sornette}},\ }\bibfield  {title} {\enquote {\bibinfo {title} {Multiplicative
  processes and power laws},}\ }\href@noop {} {\bibfield  {journal} {\bibinfo
  {journal} {Phys. Rev. E}\ }\textbf {\bibinfo {volume} {57}},\ \bibinfo
  {pages} {4811--4813} (\bibinfo {year} {1998}{\natexlab{a}})},\ \bibinfo
  {note} {https://doi.org/10.1103/PhysRevE.57.4811}\BibitemShut {NoStop}%
\bibitem [{\citenamefont {Sornette}(1998{\natexlab{b}})}]{sornette:1998PhysA}%
  \BibitemOpen
  \bibfield  {author} {\bibinfo {author} {\bibfnamefont {D.}~\bibnamefont
  {Sornette}},\ }\bibfield  {title} {\enquote {\bibinfo {title} {Linear
  stochastic dynamics with nonlinear fractal properties},}\ }\href@noop {}
  {\bibfield  {journal} {\bibinfo  {journal} {Physica A}\ }\textbf {\bibinfo
  {volume} {250}},\ \bibinfo {pages} {295--314} (\bibinfo {year}
  {1998}{\natexlab{b}})},\ \bibinfo {note}
  {https://doi.org/10.1016/S0378-4371(97)00543-8}\BibitemShut {NoStop}%
\bibitem [{\citenamefont {Levy}\ and\ \citenamefont
  {Solomon}(1996)}]{levy:1996}%
  \BibitemOpen
  \bibfield  {author} {\bibinfo {author} {\bibfnamefont {M.}~\bibnamefont
  {Levy}}\ and\ \bibinfo {author} {\bibfnamefont {S.}~\bibnamefont {Solomon}},\
  }\bibfield  {title} {\enquote {\bibinfo {title} {Power laws are logarithmic
  {B}oltzmann laws},}\ }\href@noop {} {\bibfield  {journal} {\bibinfo
  {journal} {Int. J. Mod. Phys. C}\ }\textbf {\bibinfo {volume} {7}},\ \bibinfo
  {pages} {595--601} (\bibinfo {year} {1996})},\ \bibinfo {note}
  {https://doi.org/10.1142/S0129183196000491}\BibitemShut {NoStop}%
\bibitem [{\citenamefont {Zel'dovich}\ \emph {et~al.}(1987)\citenamefont
  {Zel'dovich}, \citenamefont {Molchanov}, \citenamefont {Ruzmaikin},\ and\
  \citenamefont {Sokolov}}]{zeldovich:1987}%
  \BibitemOpen
  \bibfield  {author} {\bibinfo {author} {\bibfnamefont {Y.~B.}\ \bibnamefont
  {Zel'dovich}}, \bibinfo {author} {\bibfnamefont {S.~A.}\ \bibnamefont
  {Molchanov}}, \bibinfo {author} {\bibfnamefont {A.~A.}\ \bibnamefont
  {Ruzmaikin}}, \ and\ \bibinfo {author} {\bibfnamefont {D.~D.}\ \bibnamefont
  {Sokolov}},\ }\bibfield  {title} {\enquote {\bibinfo {title} {Intermittency
  in random media},}\ }\href@noop {} {\bibfield  {journal} {\bibinfo  {journal}
  {Sov. Phys. Usp.}\ }\textbf {\bibinfo {volume} {30}},\ \bibinfo {pages}
  {353--369} (\bibinfo {year} {1987})},\ \bibinfo {note}
  {https://doi.org/10.1070/PU1987v030n05ABEH002867}\BibitemShut {NoStop}%
\bibitem [{\citenamefont {Zanette}\ and\ \citenamefont
  {Manrubia}(1997)}]{zanette:1997}%
  \BibitemOpen
  \bibfield  {author} {\bibinfo {author} {\bibfnamefont {D.~H.}\ \bibnamefont
  {Zanette}}\ and\ \bibinfo {author} {\bibfnamefont {S.~C.}\ \bibnamefont
  {Manrubia}},\ }\bibfield  {title} {\enquote {\bibinfo {title} {Role of
  intermittency in urban development: A model of large-scale city formation},}\
  }\href@noop {} {\bibfield  {journal} {\bibinfo  {journal} {Phys. Rev. Lett.}\
  }\textbf {\bibinfo {volume} {79}},\ \bibinfo {pages} {523--526} (\bibinfo
  {year} {1997})},\ \bibinfo {note}
  {https://doi.org/10.1103/PhysRevLett.79.523}\BibitemShut {NoStop}%
\bibitem [{\citenamefont {Nirei}\ and\ \citenamefont
  {Souma}(2004)}]{nirei:2004}%
  \BibitemOpen
  \bibfield  {author} {\bibinfo {author} {\bibfnamefont {M.}~\bibnamefont
  {Nirei}}\ and\ \bibinfo {author} {\bibfnamefont {W.}~\bibnamefont {Souma}},\
  }\bibfield  {title} {\enquote {\bibinfo {title} {Income distribution and
  stochastic multiplicative process with reset event},}\ }in\ \href@noop {}
  {\emph {\bibinfo {booktitle} {The Complex Dynamics of Economic
  Interaction}}},\ \bibinfo {editor} {edited by\ \bibinfo {editor}
  {\bibfnamefont {M.}~\bibnamefont {Gallegati}}, \bibinfo {editor}
  {\bibfnamefont {A.~P.}\ \bibnamefont {Kirman}}, \ and\ \bibinfo {editor}
  {\bibfnamefont {M.}~\bibnamefont {Marsili}}}\ (\bibinfo  {publisher}
  {Springer},\ \bibinfo {address} {Berlin},\ \bibinfo {year} {2004})\ pp.\
  \bibinfo {pages} {161--168},\ \bibinfo {note}
  {https://doi.org/10.1007/978-3-642-17045-4}\BibitemShut {NoStop}%
\bibitem [{\citenamefont {Montero}, \citenamefont {Mas{\'o}-Puigdellosas},\
  and\ \citenamefont {Villarroel}(2017)}]{montero:2017}%
  \BibitemOpen
  \bibfield  {author} {\bibinfo {author} {\bibfnamefont {M.}~\bibnamefont
  {Montero}}, \bibinfo {author} {\bibfnamefont {A.}~\bibnamefont
  {Mas{\'o}-Puigdellosas}}, \ and\ \bibinfo {author} {\bibfnamefont
  {J.}~\bibnamefont {Villarroel}},\ }\bibfield  {title} {\enquote {\bibinfo
  {title} {Continuous-time random walks with reset events},}\ }\href@noop {}
  {\bibfield  {journal} {\bibinfo  {journal} {Eur. Phys. J B}\ }\textbf
  {\bibinfo {volume} {90}},\ \bibinfo {pages} {176} (\bibinfo {year} {2017})},\
  \bibinfo {note} {https://doi.org/10.1140/epjb/e2017-80348-4}\BibitemShut
  {NoStop}%
\bibitem [{\citenamefont {Evans}, \citenamefont {Majumdar},\ and\ \citenamefont
  {Schehr}(2019)}]{evans:2019}%
  \BibitemOpen
  \bibfield  {author} {\bibinfo {author} {\bibfnamefont {M.~R.}\ \bibnamefont
  {Evans}}, \bibinfo {author} {\bibfnamefont {S.~N.}\ \bibnamefont {Majumdar}},
  \ and\ \bibinfo {author} {\bibfnamefont {G.}~\bibnamefont {Schehr}},\
  }\href@noop {} {\enquote {\bibinfo {title} {Stochastic resetting and
  applications},}\ } (\bibinfo {year} {2019}),\ \bibinfo {note} {available
  online at https://arxiv.org/abs/1910.07993}\BibitemShut {NoStop}%
\bibitem [{\citenamefont {M\'ezard}, \citenamefont {Parisi},\ and\
  \citenamefont {Virasoro}(1987)}]{mezard:1987}%
  \BibitemOpen
  \bibfield  {author} {\bibinfo {author} {\bibfnamefont {M.}~\bibnamefont
  {M\'ezard}}, \bibinfo {author} {\bibfnamefont {G.}~\bibnamefont {Parisi}}, \
  and\ \bibinfo {author} {\bibfnamefont {M.}~\bibnamefont {Virasoro}},\
  }\href@noop {} {\emph {\bibinfo {title} {Spin Glass Theory and Beyond. An
  Introduction to the Replica Method and Its Applications}}}\ (\bibinfo
  {publisher} {World Scientific},\ \bibinfo {address} {Singapore},\ \bibinfo
  {year} {1987})\ \bibinfo {note} {https://doi.org/10.1142/0271}\BibitemShut
  {NoStop}%
\bibitem [{\citenamefont {Derrida}(1997)}]{derrida:1997}%
  \BibitemOpen
  \bibfield  {author} {\bibinfo {author} {\bibfnamefont {B.}~\bibnamefont
  {Derrida}},\ }\bibfield  {title} {\enquote {\bibinfo {title} {From random
  walks to spin glasses},}\ }\href@noop {} {\bibfield  {journal} {\bibinfo
  {journal} {Physica D}\ }\textbf {\bibinfo {volume} {107}},\ \bibinfo {pages}
  {186--198} (\bibinfo {year} {1997})},\ \bibinfo {note}
  {https://doi.org/10.1016/S0167-2789(97)00086-9}\BibitemShut {NoStop}%
\bibitem [{\citenamefont {Derrida}\ and\ \citenamefont
  {Jung-Muller}(1999)}]{derrida:1999}%
  \BibitemOpen
  \bibfield  {author} {\bibinfo {author} {\bibfnamefont {B.}~\bibnamefont
  {Derrida}}\ and\ \bibinfo {author} {\bibfnamefont {B.}~\bibnamefont
  {Jung-Muller}},\ }\bibfield  {title} {\enquote {\bibinfo {title} {The
  genealogical tree of a chromosome},}\ }\href@noop {} {\bibfield  {journal}
  {\bibinfo  {journal} {J. Stat. Phys.}\ }\textbf {\bibinfo {volume} {94}},\
  \bibinfo {pages} {277--298} (\bibinfo {year} {1999})},\ \bibinfo {note}
  {https://doi.org/10.1023/A:1004579800589}\BibitemShut {NoStop}%
\bibitem [{\citenamefont {Meylahn}, \citenamefont {Sabhapandit},\ and\
  \citenamefont {Touchette}(2015)}]{meilahn:2015}%
  \BibitemOpen
  \bibfield  {author} {\bibinfo {author} {\bibfnamefont {J.~M.}\ \bibnamefont
  {Meylahn}}, \bibinfo {author} {\bibfnamefont {S.}~\bibnamefont
  {Sabhapandit}}, \ and\ \bibinfo {author} {\bibfnamefont {H.}~\bibnamefont
  {Touchette}},\ }\bibfield  {title} {\enquote {\bibinfo {title} {Large
  deviations for {M}arkov processes with resetting},}\ }\href@noop {}
  {\bibfield  {journal} {\bibinfo  {journal} {Phys. Rev. E}\ }\textbf {\bibinfo
  {volume} {92}},\ \bibinfo {pages} {062148} (\bibinfo {year} {2015})},\
  \bibinfo {note} {https://doi.org/10.1103/PhysRevE.92.062148}\BibitemShut
  {NoStop}%
\bibitem [{\citenamefont {Gabaix}(2009)}]{gabaix:2009}%
  \BibitemOpen
  \bibfield  {author} {\bibinfo {author} {\bibfnamefont {X.}~\bibnamefont
  {Gabaix}},\ }\bibfield  {title} {\enquote {\bibinfo {title} {Power laws in
  economics and finance},}\ }\href@noop {} {\bibfield  {journal} {\bibinfo
  {journal} {Ann. Rev. Econ.}\ }\textbf {\bibinfo {volume} {1}},\ \bibinfo
  {pages} {255--293} (\bibinfo {year} {2009})},\ \bibinfo {note}
  {https://doi.org/10.3386/w14299}\BibitemShut {NoStop}%
\bibitem [{\citenamefont {Gibrat}(1931)}]{gibrat:1931}%
  \BibitemOpen
  \bibfield  {author} {\bibinfo {author} {\bibfnamefont {R.}~\bibnamefont
  {Gibrat}},\ }\href@noop {} {\emph {\bibinfo {title} {Les in\'egalit\'es
  \'economiques}}}\ (\bibinfo  {publisher} {Recueil Sirey},\ \bibinfo {address}
  {Paris},\ \bibinfo {year} {1931})\BibitemShut {NoStop}%
\bibitem [{\citenamefont {Lagi}\ \emph {et~al.}(2011)\citenamefont {Lagi},
  \citenamefont {Bar-Yam}, \citenamefont {Bertrand},\ and\ \citenamefont
  {Bar-Yam}}]{lagi:2011}%
  \BibitemOpen
  \bibfield  {author} {\bibinfo {author} {\bibfnamefont {M.}~\bibnamefont
  {Lagi}}, \bibinfo {author} {\bibfnamefont {Y.}~\bibnamefont {Bar-Yam}},
  \bibinfo {author} {\bibfnamefont {K.~Z.}\ \bibnamefont {Bertrand}}, \ and\
  \bibinfo {author} {\bibfnamefont {Y.}~\bibnamefont {Bar-Yam}},\ }\bibfield
  {title} {\enquote {\bibinfo {title} {The food crises: A quantitative model of
  food prices including speculators and ethanol conversion},}\ }\href@noop {}
  {\bibfield  {journal} {\bibinfo  {journal} {SSRN}\ ,\ \bibinfo {pages}
  {1932247}} (\bibinfo {year} {2011})},\ \bibinfo {note}
  {https://doi.org/10.2139/ssrn.1932247}\BibitemShut {NoStop}%
\bibitem [{\citenamefont {Caprio}\ and\ \citenamefont
  {Klingebiel}(1999)}]{caprio:1999}%
  \BibitemOpen
  \bibfield  {author} {\bibinfo {author} {\bibfnamefont {G.}~\bibnamefont
  {Caprio}}\ and\ \bibinfo {author} {\bibfnamefont {D.}~\bibnamefont
  {Klingebiel}},\ }\href@noop {} {\emph {\bibinfo {title} {Bank Insolvencies:
  Cross-Country Experience}}}\ (\bibinfo  {publisher} {The World Bank},\
  \bibinfo {address} {Washington, DC},\ \bibinfo {year} {1999})\ \bibinfo
  {note} {https://doi.org/10.1596/1813-9450-1620}\BibitemShut {NoStop}%
\bibitem [{\citenamefont {Taleb}(2007{\natexlab{b}})}]{taleb:2007AS}%
  \BibitemOpen
  \bibfield  {author} {\bibinfo {author} {\bibfnamefont {N.~N.}\ \bibnamefont
  {Taleb}},\ }\bibfield  {title} {\enquote {\bibinfo {title} {Black swans and
  the domains of statistics},}\ }\href@noop {} {\bibfield  {journal} {\bibinfo
  {journal} {Am. Stat.}\ }\textbf {\bibinfo {volume} {61}},\ \bibinfo {pages}
  {198--200} (\bibinfo {year} {2007}{\natexlab{b}})},\ \bibinfo {note}
  {https://doi.org/10.1198/000313007X219996}\BibitemShut {NoStop}%
\bibitem [{\citenamefont {Montero}\ and\ \citenamefont
  {Villarroel}(2016)}]{montero:2016}%
  \BibitemOpen
  \bibfield  {author} {\bibinfo {author} {\bibfnamefont {M.}~\bibnamefont
  {Montero}}\ and\ \bibinfo {author} {\bibfnamefont {J.}~\bibnamefont
  {Villarroel}},\ }\bibfield  {title} {\enquote {\bibinfo {title} {Directed
  random walk with random restarts: The {S}isyphus random walk},}\ }\href@noop
  {} {\bibfield  {journal} {\bibinfo  {journal} {Phys. Rev. E}\ }\textbf
  {\bibinfo {volume} {94}},\ \bibinfo {pages} {032132} (\bibinfo {year}
  {2016})},\ \bibinfo {note}
  {https://doi.org/10.1103/PhysRevE.94.032132}\BibitemShut {NoStop}%
\bibitem [{\citenamefont {Heubach}\ and\ \citenamefont
  {Mansour}(2009)}]{compos}%
  \BibitemOpen
  \bibfield  {author} {\bibinfo {author} {\bibfnamefont {S.}~\bibnamefont
  {Heubach}}\ and\ \bibinfo {author} {\bibfnamefont {T.}~\bibnamefont
  {Mansour}},\ }\href@noop {} {\emph {\bibinfo {title} {Combinatorics of
  Compositions and Words. Discrete Mathematics and its Applications}}}\
  (\bibinfo  {publisher} {CRC Press},\ \bibinfo {address} {Boca Raton},\
  \bibinfo {year} {2009})\BibitemShut {NoStop}%
\bibitem [{\citenamefont {Redner}(1990)}]{redner}%
  \BibitemOpen
  \bibfield  {author} {\bibinfo {author} {\bibfnamefont {S.}~\bibnamefont
  {Redner}},\ }\bibfield  {title} {\enquote {\bibinfo {title} {Random
  multiplicative processes: An elementary tutorial},}\ }\href@noop {}
  {\bibfield  {journal} {\bibinfo  {journal} {Am. J. Phys.}\ }\textbf {\bibinfo
  {volume} {58}},\ \bibinfo {pages} {267--273} (\bibinfo {year} {1990})},\
  \bibinfo {note} {https://doi.org/10.1119/1.16497}\BibitemShut {NoStop}%
\bibitem [{\citenamefont {Taleb}(2009)}]{taleb:2008}%
  \BibitemOpen
  \bibfield  {author} {\bibinfo {author} {\bibfnamefont {N.~N.}\ \bibnamefont
  {Taleb}},\ }\bibfield  {title} {\enquote {\bibinfo {title} {Finiteness of
  variance is irrelevant in the practice of quantitative finance},}\
  }\href@noop {} {\bibfield  {journal} {\bibinfo  {journal} {Complexity}\
  }\textbf {\bibinfo {volume} {14}},\ \bibinfo {pages} {66--76} (\bibinfo
  {year} {2009})},\ \bibinfo {note}
  {https://doi.org/10.1002/cplx.20263}\BibitemShut {NoStop}%
\bibitem [{\citenamefont {Brooks}(2019)}]{boot}%
  \BibitemOpen
  \bibfield  {author} {\bibinfo {author} {\bibfnamefont {C.}~\bibnamefont
  {Brooks}},\ }\href@noop {} {\emph {\bibinfo {title} {Introductory
  Econometrics for Finance}}}\ (\bibinfo  {publisher} {Cambridge University
  Press},\ \bibinfo {address} {Oxford},\ \bibinfo {year} {2019})\ \bibinfo
  {note} {https://doi.org/10.1017/9781108524872}\BibitemShut {NoStop}%
\bibitem [{\citenamefont {Novak}(2011)}]{rare}%
  \BibitemOpen
  \bibfield  {author} {\bibinfo {author} {\bibfnamefont {S.~Y.}\ \bibnamefont
  {Novak}},\ }\href@noop {} {\emph {\bibinfo {title} {Extreme Value Methods
  with Applications to Finance}}}\ (\bibinfo  {publisher} {Chapman \& Hall/CRC
  Press},\ \bibinfo {address} {London},\ \bibinfo {year} {2011})\BibitemShut
  {NoStop}%
\bibitem [{\citenamefont {Taleb}(2005)}]{taleb:2005}%
  \BibitemOpen
  \bibfield  {author} {\bibinfo {author} {\bibfnamefont {N.~N.}\ \bibnamefont
  {Taleb}},\ }\href@noop {} {\emph {\bibinfo {title} {Fooled by Randomness. The
  Hidden Role of Chance in Life and in the Markets}}}\ (\bibinfo  {publisher}
  {Random House},\ \bibinfo {address} {New York},\ \bibinfo {year}
  {2005})\BibitemShut {NoStop}%
\bibitem [{\citenamefont {Ku\'smierz}\ and\ \citenamefont
  {Toyoizumi}(2019)}]{kusmierz:2019}%
  \BibitemOpen
  \bibfield  {author} {\bibinfo {author} {\bibfnamefont {{\L{}}.}~\bibnamefont
  {Ku\'smierz}}\ and\ \bibinfo {author} {\bibfnamefont {T.}~\bibnamefont
  {Toyoizumi}},\ }\bibfield  {title} {\enquote {\bibinfo {title} {Robust random
  search with scale-free stochastic resetting},}\ }\href@noop {} {\bibfield
  {journal} {\bibinfo  {journal} {Phys. Rev. E}\ }\textbf {\bibinfo {volume}
  {100}},\ \bibinfo {pages} {032110} (\bibinfo {year} {2019})},\ \bibinfo
  {note} {https://doi.org/10.1103/PhysRevE.100.032110}\BibitemShut {NoStop}%
\bibitem [{\citenamefont {Beare}\ and\ \citenamefont
  {Toda}(2019)}]{beare:2019}%
  \BibitemOpen
  \bibfield  {author} {\bibinfo {author} {\bibfnamefont {B.~K.}\ \bibnamefont
  {Beare}}\ and\ \bibinfo {author} {\bibfnamefont {A.~A.}\ \bibnamefont
  {Toda}},\ }\href@noop {} {\enquote {\bibinfo {title} {Geometrically stopped
  {Markovian} random growth processes and {Pareto} tails},}\ } (\bibinfo {year}
  {2019}),\ \bibinfo {note} {available online at
  https://arxiv.org/abs/1712.01431v2}\BibitemShut {NoStop}%
\bibitem [{\citenamefont {Buend\'{\i}a}, \citenamefont {Mu\~noz},\ and\
  \citenamefont {Manrubia}(2018)}]{buendia:2018}%
  \BibitemOpen
  \bibfield  {author} {\bibinfo {author} {\bibfnamefont {V.}~\bibnamefont
  {Buend\'{\i}a}}, \bibinfo {author} {\bibfnamefont {M.~A.}\ \bibnamefont
  {Mu\~noz}}, \ and\ \bibinfo {author} {\bibfnamefont {S.}~\bibnamefont
  {Manrubia}},\ }\bibfield  {title} {\enquote {\bibinfo {title} {Limited role
  of spatial self-structuring in emergent trade-offs during pathogen
  evolution},}\ }\href@noop {} {\bibfield  {journal} {\bibinfo  {journal} {Sci.
  Rep.}\ }\textbf {\bibinfo {volume} {8}},\ \bibinfo {pages} {12476} (\bibinfo
  {year} {2018})},\ \bibinfo {note}
  {https://doi.org/10.1038/s41598-018-30945-1}\BibitemShut {NoStop}%
\bibitem [{\citenamefont {Adler}(1993)}]{adler:1993}%
  \BibitemOpen
  \bibfield  {author} {\bibinfo {author} {\bibfnamefont {F.~R.}\ \bibnamefont
  {Adler}},\ }\bibfield  {title} {\enquote {\bibinfo {title} {Migration alone
  can produce persistence of host-parasitoid models},}\ }\href@noop {}
  {\bibfield  {journal} {\bibinfo  {journal} {Am. Nat.}\ }\textbf {\bibinfo
  {volume} {141}},\ \bibinfo {pages} {642--650} (\bibinfo {year} {1993})},\
  \bibinfo {note} {https://doi.org/10.1086/285496}\BibitemShut {NoStop}%
\bibitem [{\citenamefont {Yamamoto}\ and\ \citenamefont
  {Yamazaki}(2012)}]{yamamoto:2012}%
  \BibitemOpen
  \bibfield  {author} {\bibinfo {author} {\bibfnamefont {K.}~\bibnamefont
  {Yamamoto}}\ and\ \bibinfo {author} {\bibfnamefont {Y.}~\bibnamefont
  {Yamazaki}},\ }\bibfield  {title} {\enquote {\bibinfo {title} {Power-law
  behavior in a cascade process with stopping events: A solvable model},}\
  }\href@noop {} {\bibfield  {journal} {\bibinfo  {journal} {Phys. Rev. E}\
  }\textbf {\bibinfo {volume} {85}},\ \bibinfo {pages} {011145} (\bibinfo
  {year} {2012})},\ \bibinfo {note}
  {https://doi.org/10.1103/PhysRevE.85.011145}\BibitemShut {NoStop}%
\end{thebibliography}%

\end{document}